\documentclass[12pt]{article}
\pdfoutput=1
\usepackage[pdftex]{graphics}
\usepackage{graphicx}
\usepackage{hyperref}
\usepackage[square, comma, numbers, sort&compress]{natbib}
\usepackage{caption,subfigure}
\subcapfont{\bf}
\renewcommand{\figurename}{{\bf Figure}} 
\makeatletter
\renewcommand{\fnum@figure}[1]{\textbf{\figurename~\thefigure}:}

\makeatother 

\newcommand\pubnumber{SLAC--PUB--14433 \\ SU--ITP--11/20}
\newcommand\pubdate{}%?????, 2011}

\def\SLAC{SLAC, Menlo Park, CA 94025 \\ SITP, Stanford University,
Stanford, CA 94305}

\def\emailone{\footnote{janko@stanford.edu}}                     
\def\emailtwo{\footnote{larkoski@stanford.edu}}           

\def\Title#1{\begin{center} {\Large #1 } \end{center}}
\def\Author#1{\begin{center}{ \sc #1} \end{center}}
\def\Address#1{\begin{center}{ \it #1} \end{center}}

\newcommand\pubblock{\rightline{\begin{tabular}{l} \pubnumber\\
         \pubdate \end{tabular}}}
\newenvironment{Abstract}{\begin{quotation} \begin{center}
                       ABSTRACT
     \end{center}\bigskip  }{\end{quotation}}

\def\Acknowledgements{\bigskip  \bigskip \begin{center} \begin{large}
             \bf ACKNOWLEDGEMENTS \end{large}\end{center}}
\def\beq{\begin{equation}}
\def\eeq#1{\label{#1}\end{equation}}
\def\eeqn{\end{equation}}

\newenvironment{Eqnarray}%
   {\arraycolsep 0.14em\begin{eqnarray}}{\end{eqnarray}}
\def\beqa{\begin{Eqnarray}}
\def\eeqa#1{\label{#1}\end{Eqnarray}}
\def\eeqan{\end{Eqnarray}}

\def\leqn#1{(\ref{#1})}

\let\bar=\overbar

\def\lsim{\mathrel{\raise.3ex\hbox{$<$\kern-.75em\lower1ex\hbox{$\sim$}}}}
\def\gsim{\mathrel{\raise.3ex\hbox{$>$\kern-.75em\lower1ex\hbox{$\sim$}}}}

\def\del{\partial}
\def\Dslash{\not{\hbox{\kern-4pt $D$}}}
\def\dslash{\not{\hbox{\kern-2pt $\del$}}}
\def\pslash{\not{\hbox{\kern-2pt $p$}}}
\def\qslash{\not{\hbox{\kern-2pt $q$}}}
\def\kslash{\not{\hbox{\kern-2pt $k$}}}
\def\oneslash{\not{\hbox{\kern-2pt $1$}}}
\def\twoslash{\not{\hbox{\kern-2pt $2$}}}
\def\threeslash{\not{\hbox{\kern-2pt $3$}}}
\def\fourslash{\not{\hbox{\kern-2pt $4$}}}
\def\aslash{\not{\hbox{\kern-2pt $a$}}}
\def\rslash{\not{\hbox{\kern-2pt $r$}}}
\def\Jslash{\not{\hbox{\kern-2pt $J$}}}

\def\msb{{\bar{\scriptsize M \kern -1pt S}}}

\def\drb{{\bar{\scriptsize D \kern -1pt R}}}

\def\apb#1 {  \langle #1 ] }

\makeatletter
\def\section{\@startsection{section}{0}{\z@}{5.5ex plus .5ex minus
 1.5ex}{2.3ex plus .2ex}{\large\bf}}
\def\subsection{\@startsection{subsection}{1}{\z@}{3.5ex plus .5ex minus
 1.5ex}{1.3ex plus .2ex}{\normalsize\bf}}
\def\subsubsection{\@startsection{subsubsection}{2}{\z@}{-3.5ex plus
-1ex minus  -.2ex}{2.3ex plus .2ex}{\normalsize\sl}}

\renewcommand{\@makecaption}[2]{%
   \vskip 10pt
   \setbox\@tempboxa\hbox{\small #1: #2}
   \ifdim \wd\@tempboxa >\hsize     % IF longer than one line:
       \small #1: #2\par          %   THEN set as ordinary paragraph.
     \else                        %   ELSE  center.
       \hbox to\hsize{\hfil\box\@tempboxa\hfil}
   \fi}

\def\apb#1#2#3{\langle #1  #2  #3 ]}

\begin{document}
\begin{titlepage}
\pubblock

\vfill
\Title{Jet Substructure Without Trees}
\vfill
\Author{Martin Jankowiak\emailone and Andrew J. Larkoski\emailtwo}
\Address{\SLAC}
\vfill
\begin{Abstract}

We present an alternative approach to identifying and characterizing jet substructure.  
An angular correlation function is introduced that can be used to extract 
angular and mass scales within a jet without reference to a clustering
algorithm.  This procedure gives rise to a number of useful jet observables.
As an application, we construct a top quark tagging algorithm that
is competitive with existing methods.  

\end{Abstract}
\vfill
\vfill
\end{titlepage}
\def\thefootnote{\fnsymbol{footnote}}
\setcounter{footnote}{1}
%
%\tableofcontents
\newpage
\setcounter{page}{1}

\section{Introduction}

In preparation for the LHC, the past several years have seen 
extensive work on various aspects of collider searches. 
With the excellent resolution of the ATLAS and CMS detectors 
as a catalyst, one area that has undergone significant development 
is jet substructure physics.  The use of jet substructure techniques, which probe 
the fine-grained details of how energy is distributed in jets, has two broad 
goals.  First, measuring more than just the bulk properties of jets 
allows for additional probes of QCD.  For example, jet substructure 
measurements can be compared against precision perturbative 
QCD calculations or used to tune Monte Carlo
event generators.  Second, jet substructure allows for additional 
handles in event discrimination.  These handles could play an 
important role at the LHC in discriminating between signal and 
background events in a wide variety of particle searches.  For example, 
Monte Carlo studies indicate that jet substructure 
techniques allow for efficient reconstruction of boosted heavy 
objects such as the $W^\pm$ and $Z^0$ gauge bosons \citep{Seymour:1993mx,Butterworth:2002tt,tweedie,Cui:2010km}, 
the top quark \citep{sterman,thalerwang,hep,stoprecon,brooijmans,toptag}, 
and the Higgs boson \citep{bdrs,Kribs:2009yh,kim,Chen:2010wk,Falkowski:2010hi,Hackstein:2010wk}.  

At least two broad classes of jet substructure 
techniques have been developed.  The first class employs jet
shape observables to probe energy distribution in jets.  
The second class makes use of the clustering tree of a jet as 
constructed by the Cambridge-Aachen (CA) \citep{CAalg}
or $k_T$ \citep{ktalg} sequential jet clustering algorithms to identify and characterize subjets within the jet.

Jet shape observables offer a measure of how energy is 
distributed within a jet.  The energy distribution of a jet
is determined by a variety of factors, including 
heavy particle decays, color flow, 
and the dynamics of the parton shower.
Different jet shape observables have been constructed to quantify these 
\citep{template,subjetty,Almeida:2008yp,Gallicchio:2010sw,Hook:2011cq,Soper:2011cr}
and other aspects of jet substructure.  Infrared
and collinear (IRC) safe observables can in principle be computed in perturbation theory
or modeled with Monte Carlo simulations and then compared to 
experimental results.  Combining different jet shape observables
has been shown to provide for effective discrimination in a variety
of different scenarios (see e.g.~\citep{tmva}).  
A disadvantage of jet shape observables is that, because they can only
be computed once the constituents of the jet have been defined, they
cannot be used to determine how to most effectively select jets within a given event.
In particular a jet shape observable is only as good as the choice of particles
that define the jet. As a result jet shape observables do not offer
a way of selectively removing likely contamination from underlying event or pile-up\footnote{See however \citep{Alon:2011xb}.}.

The CA and $k_T$ sequential jet algorithms are defined by metrics
$d_{ij}$ that have been chosen with the goal of constructing clustering trees that closely approximate
the perturbative QCD parton shower.  
The first few branches of the clustering tree can be used to decompose
a jet into subjets.  This unclustering
procedure has seen a wide variety of phenomenological 
applications, especially in the context of tagging jets
that result from boosted heavy particle decays, ${\it e.g.}\;$filtering in
boosted Higgs searches \citep{bdrs}.
A closely related procedure, referred to as pruning \citep{EllisWalsh}, vetoes
on QCD-like branches with the goal of sharpening jet mass resolution.
This family of procedures
offers a number of tunable parameters, allowing the user to control
how much and what kind of substructure is identified.
A disadvantage of these procedures is that, in order for them
to be most effective, the clustering tree must accurately 
reconstruct the parton shower history of the jet.  In practice
the CA and $k_T$ algorithms reconstruct the most probable shower
history, which need not coincide with the actual shower history.
In addition, the parameters which define the 
unclustering typically impose a hard line
between QCD-like behavior and non-QCD-like behavior that 
can fail to accommodate jets that deviate too much from ``most probable'' jets.

The goal of this paper is to explore an alternative procedure for
identifying and characterizing substructure within jets.  The discussion
is organized as follows.  
In Section 2, we introduce the ``angular correlation function'' $\mathcal{G}(R)$ and discuss how
structure in $\mathcal{G}(R)$ can be used to construct
IRC safe jet observables.
    In particular we use $\mathcal{G}(R)$ 
to extract angular scales $R_*$ and mass scales $m_*$ directly from the constituents of a jet without use of a clustering tree.
These angular and mass scales correspond to the angular separations and invariant masses of pairs of
hard substructure in the jet.
In Section 3, we present an application of these
ideas to the tagging of boosted top quarks.  We find
that the resulting top tagging algorithm is competitive with other methods in the literature.
Given the straightforward approach we take in applying $\mathcal{G}(R)$ to top tagging, this good performance `out of the box' is encouraging.
In Section 4 we discuss other possible applications
of the methods introduced in this paper.    

\section{Angular Correlation Function}

To characterize substructure in a jet $J$ we define the angular correlation function $\mathcal{G}(R)$ as
\beq
\mathcal{G} (R)\equiv\frac{\sum\limits_{i\ne j}p_{T i}p_{T j}
\Delta R_{ij}^2\Theta (R-\Delta R_{ij})}
{\sum\limits_{i \ne j}p_{T i}p_{T j}\Delta R_{ij}^2}
 \approx \frac{\sum\limits_{i\ne j}p_i \!\cdot\! p_j \Theta (R-\Delta R_{ij})}
{\sum\limits_{i \ne j}p_i \!\cdot\! p_j} 
\eeq{corrfunc}
where the sum runs over all pairs of constituents of $J$ and $\Theta(x)$ is the Heaviside step function.  Here $p_{T i}$ is the
transverse momentum of constituent $i$, and $\Delta R_{ij}$
is the Euclidean distance between
$i$ and $j$ in the pseudorapidity ($\eta$) and azimuthal angle
 ($\phi$) plane: $\Delta R_{ij}^2=(\eta_i-\eta_j)^2+(\phi_i-\phi_j)^2.$ 
\begin{figure}
\centering
    \includegraphics[width=9.1cm]{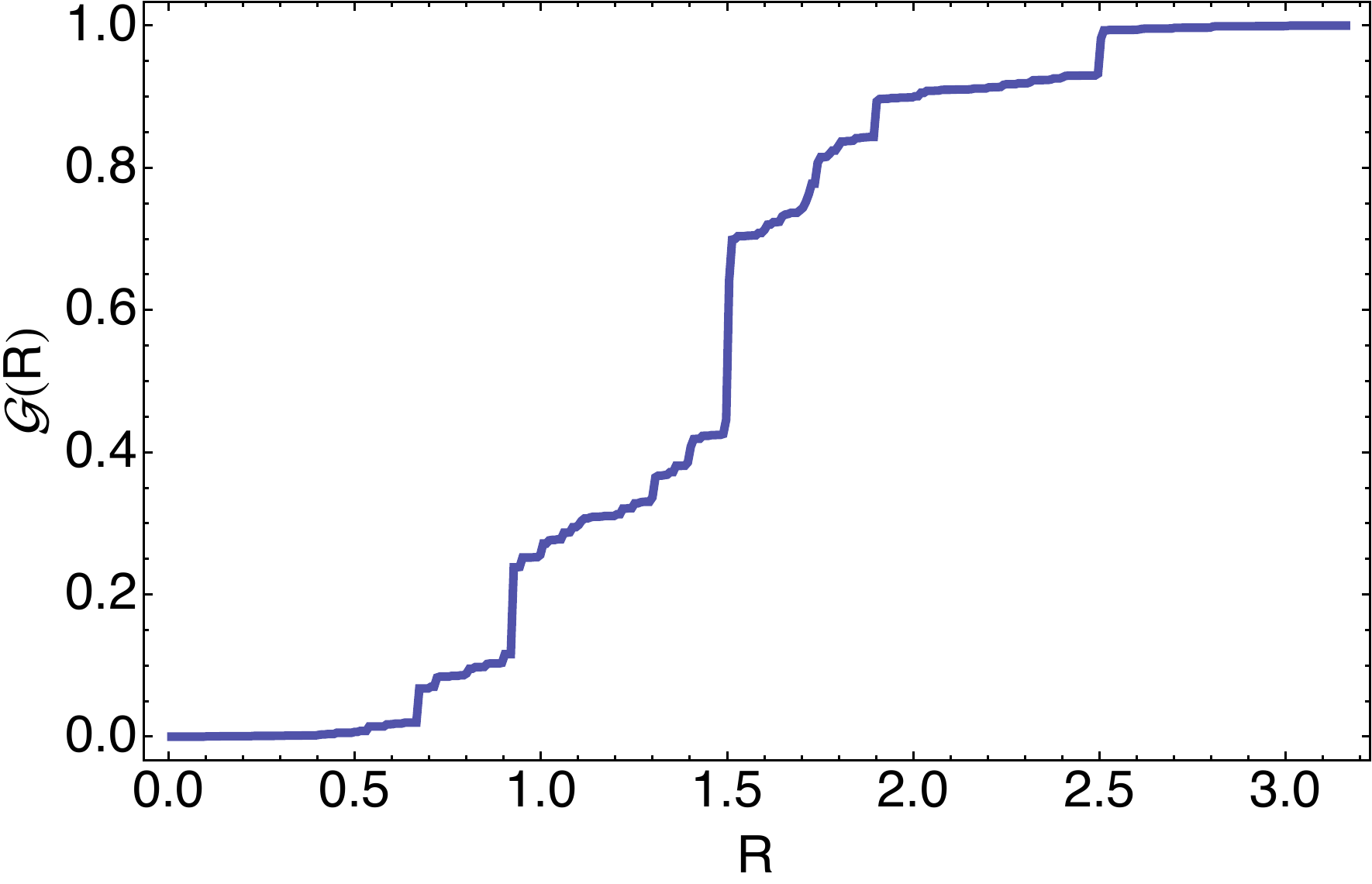}
\caption{The angular correlation function $\mathcal{G}(R)$ for a sample top jet.}\label{gcurvetop}
\end{figure}
On the LHS of Eq.~\leqn{corrfunc} the dependence
on transverse momenta is fixed by collinear safety.
Provided that  $\Delta R_{ij}$ is raised to a positive power, the entire expression is IRC safe.
  We choose
$\Delta R_{ij}^2$ in Eq.~\leqn{corrfunc} so that $\mathcal{G}(R)$ 
has a clear physical interpretation:
$\mathcal{G}(R)$ is the (fractional) mass contribution from constituents
separated by an angular distance of $R$ or less.  An important point here is that $R$ does {\it not}
mark the distance with respect to any fixed center.

For a jet with no substructure, $\mathcal{G}(R)$ is featureless.  In contrast,
if a jet has significant substructure at an angular scale $R=R_*$, $\mathcal{G}(R)$ 
exhibits a discontinuous cliff
at $R=R_*$, see Fig.$\;\ref{gcurvetop}$.  Such a cliff corresponds
 to two or more hard subjets separated by
a distance $R_*$ from one another, with the cliff height determined by
the invariant mass of the subjets. Notice that these cliffs are closely related to mass drops
 as exploited in a variety of jet substructure studies \citep{bdrs,toptag,hep,stoprecon,Kribs:2009yh}. We expect
that a typical QCD jet will have an angular correlation function that is more or less smoothly varying
without any sharp cliffs, while for a jet with significant substructure 
$\mathcal{G}(R)$ will have one or more sharp cliffs at angular scales
$R=R_*$ corresponding to distinct separations between
 hard subjets in the jet.   This suggests several jet
observables that can be defined from $\mathcal{G}(R)$.  Given a procedure
for finding cliffs in $\mathcal{G}(R)$, we can consider: (i) the total number of cliffs;
(ii) the angular scales $R=R_*$ at which cliffs are found; and (iii) the cliff heights
 at each $R=R_*$.  We 
will see that, once suitably defined, each of the resulting observables proves
useful in characterizing substructure within jets.  

In effect, $\mathcal{G}(R)$ defines a continuous family of jet shape observables.  
Each $\mathcal{G}(R_0)$
for a given $R_0$ differs from most jet shape observables in that:
(i) it does not contain any preferred or reference four-vectors (e.g.~the energy center of the jet);
and (ii) it involves a sum over {\it two-particle} correlations.
For example, the radial jet energy profile $\psi(R)$ as in \citep{Abe:1992wv,Ellis:1992qq}  
quantifies the fraction of a jet's energy that is contained within an angular distance $R$ of the
center of the jet.  Although $\psi(R)$ for a top jet will exhibit discontinuous cliffs at particular
angular scales, these scales are not useful for characterizing the substructure of the jet.  This is
because the resulting angular scales, which are defined with respect to the jet center, cannot be
used to reconstruct the separations between the three top subjets.   In addition, the invariant masses
of pairs of subjets are not accessible from $\psi(R)$.  
The angular correlation function $\mathcal{G}(R)$ 
is closer in spirit to factorial moments as in \citep{factorial}, which were introduced to quantify scaling behavior in multi-particle production.

\begin{figure}[t]
\centering
\subfigure[]
{
    \includegraphics[width=7.2cm]{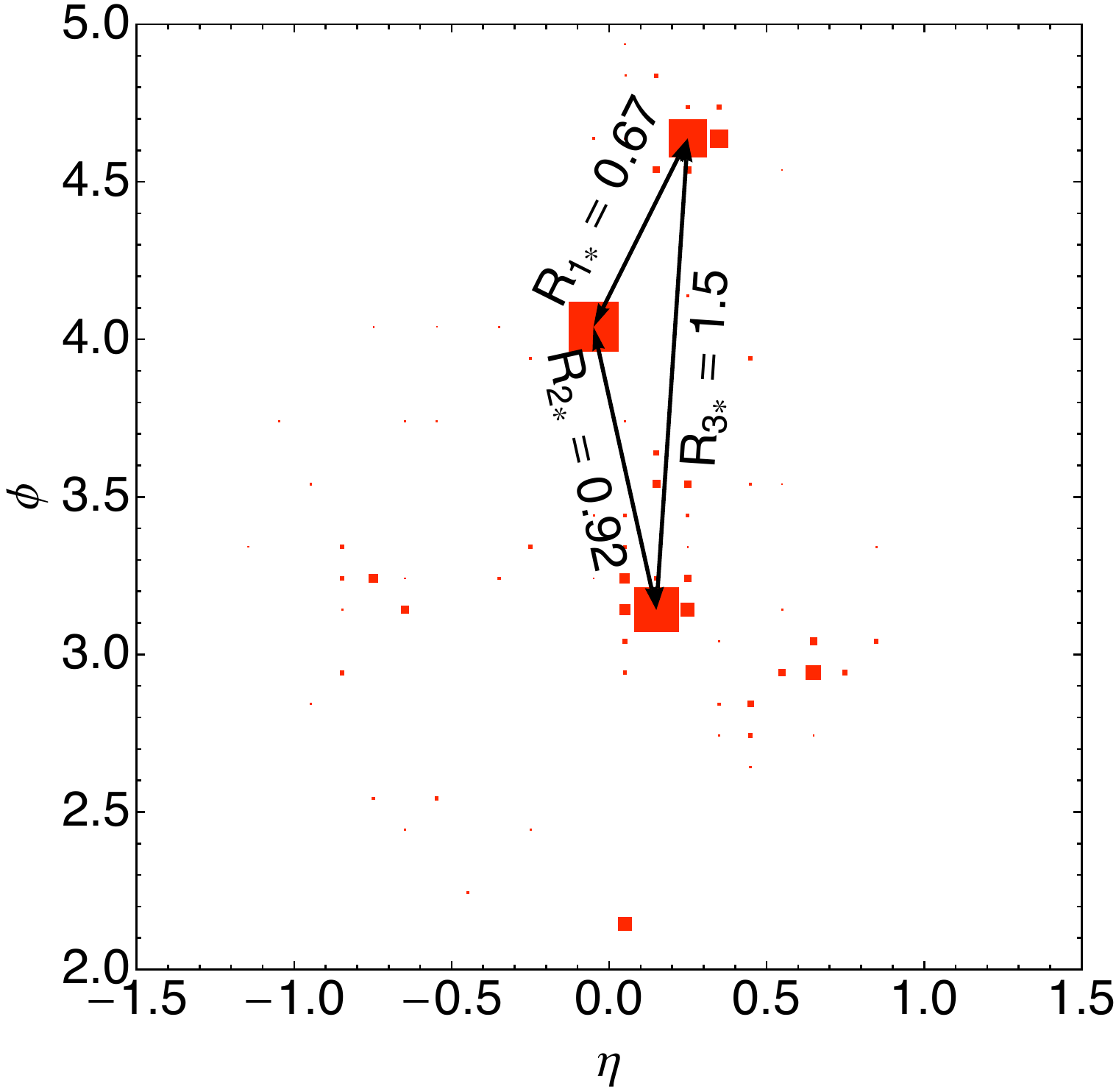}\label{legotop}
}
\hspace{.1cm}
\subfigure[]
{
    \includegraphics[width=7.0cm]{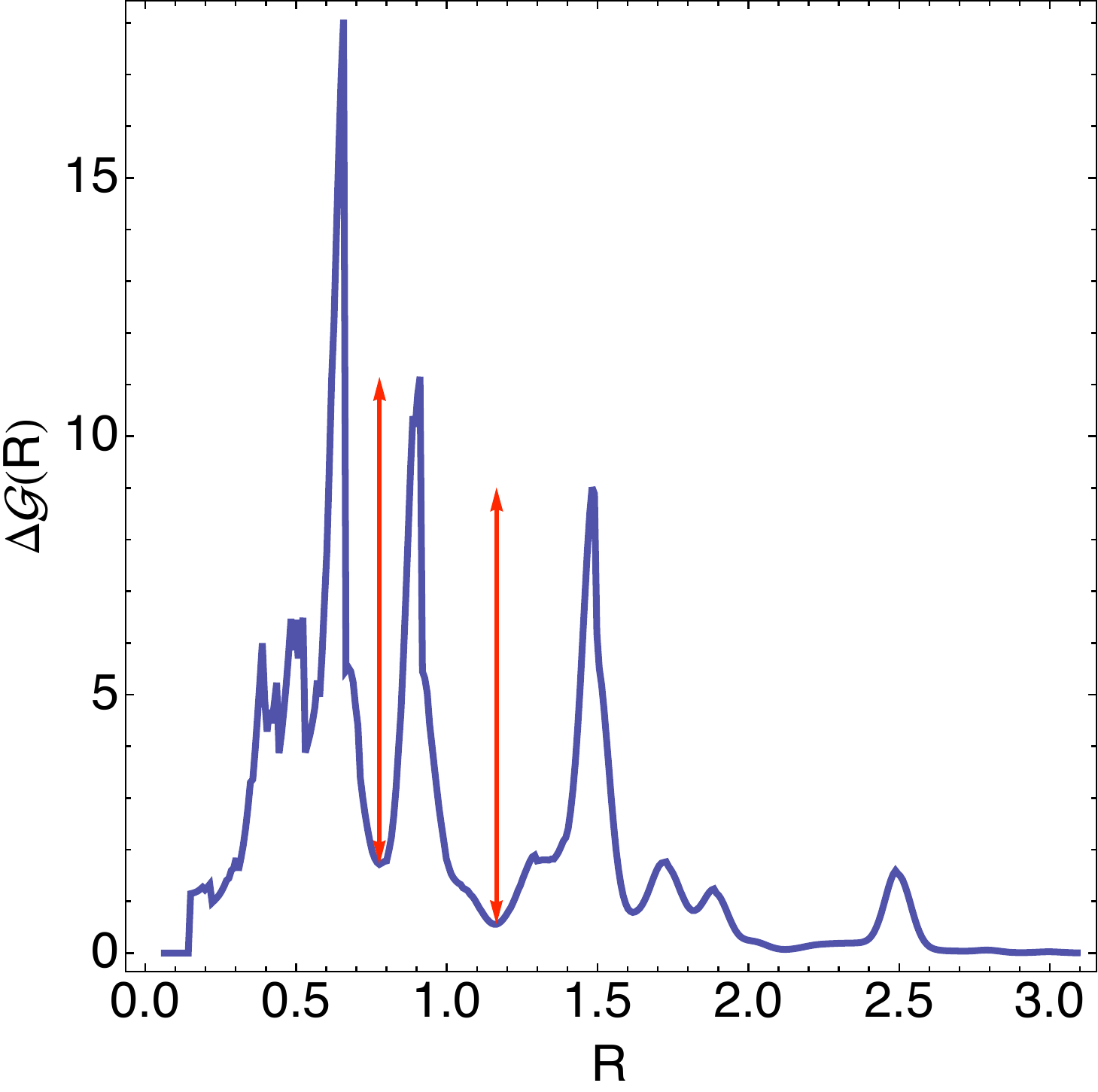}\label{masspeakplottop}
}
\caption{$p_T$ plot and angular structure function $\Delta \mathcal{G}(R)$ for the top jet whose $\mathcal{G} (R)$ is illustrated in 
Fig.~\ref{gcurvetop}.  
{\bf (a)}  The $p_T$ plot depicts the transverse energy deposited in calorimeter cells of size $0.1\times 0.1$ in $(\eta,\phi)$ with the
area of each red square proportional to the $p_T$.  This top has $p_T\sim 300$ GeV and a clean three-pronged substructure. {\bf (b)} 
For a minimum prominence of 4.0, $\Delta \mathcal{G} (R)$ has three peaks with 
$R_{1*}=0.66$, $R_{2*}=0.91$, and $R_{3*}=1.48$.  The red arrows illustrate the prominence of 
the two peaks at $R_{2*}$ and $R_{3*}$.}
\label{topjet}
\end{figure}

In order for the observables derived from $\mathcal{G}(R)$ to be useful, care must be taken in defining them.  We find
that, instead of directly finding cliffs in $\mathcal{G}(R)$, it is preferable to find
peaks in a suitably chosen derivative of $\mathcal{G}(R)$.  In particular, because
we are interested in ratios of mass scales, we should look for structure in $\log\mathcal{G}(R)$\footnote{
The normalization in $\mathcal{G}(R)$ has been chosen with this logarithm in mind: $\mathcal{G}(R)$ increases 
monotonically from 0 to 1 as $R$ increases from $R= 0$ to $R = \max \Delta R_{ij}$.}.
Because QCD is approximately scale invariant, structure in $\log\mathcal{G}(R)$ should
be identified by calculating derivatives with respect to $\log R$.  
Since ${d / d \log R}= R \ {d / d R} $,
this choice ensures that noise in $\log\mathcal{G}(R)$ 
at small $R$ does not result in extraneous peaks.
This suggests that the quantity
of interest is $d\log\mathcal{G}(R)/d\log R$.  A concern with $d\log\mathcal{G}(R)/d\log R$ is
that the derivative produces a delta function $\delta (R-\Delta R_{ij})$; as
a consequence, $d\log\mathcal{G}(R)/d\log R$ defines a noisy function of $R$. 
 Therefore, to identify
structure in $\log\mathcal{G}(R)$ we define an ``angular structure function'' $\Delta \mathcal{G}(R)$
by replacing the delta function in $d\log\mathcal{G}(R)/d\log R$ with a smooth kernel $K(x)$: 
\beq
\Delta \mathcal{G} (R)\equiv R \ \frac{\sum\limits_{i\ne j}p_{T i}p_{T j}\Delta R_{ij}^2  
K(R-\Delta R_{ij})}
{\sum\limits_{i \ne j}p_{T i}p_{T j}\Delta R_{ij}^2 \Theta(R-\Delta R_{ij})}
\eeq{masspeak}
In the following we choose a gaussian 
$K(x)=e^{-x^2/{dR}^2}/\sqrt{\pi dR^2}$ with $dR=0.06$.  We find
that this choice reduces noise substantially.  This value of $dR$ was selected after scanning a range
$dR \in [0.02, 0.12]$ and choosing $dR$ to maximize the performance 
of the top tagging algorithm presented in Sec.~3.

\begin{figure}[t]
\centering
    \includegraphics[width=6.1cm]{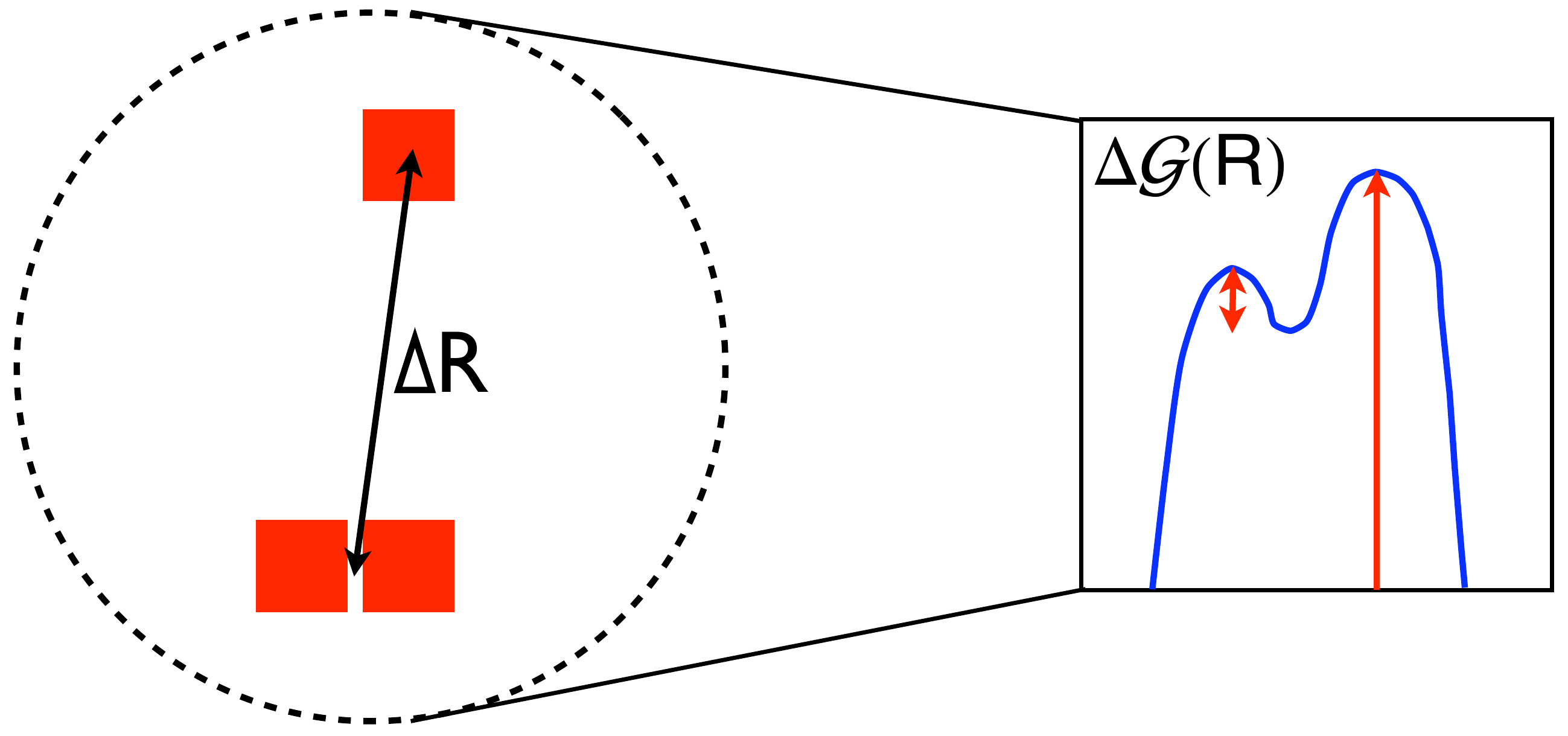}
\caption{An illustration of how prominence requirements, by selecting peaks that stand out above background noise, prevent angular scales from being double-counted.}\label{zoomin}
\end{figure}

To identify angular scales $R=R_*$ in the jet that correspond to distinct hard substructure in the event, it is important to 
find peaks in $\Delta \mathcal{G}(R)$ in a way that is robust against noise.\footnote{Using the kernel $K(x)$ 
reduces the noise in $\Delta \mathcal{G}(R)$ but does not do so completely.}  
For this purpose we borrow a concept from geography called (topographic) prominence \citep{prom}.  
The prominence of the highest peak 
is defined as its height.  In the
mountaineering analogy, the prominence
of any lower peak $P$ is defined as the minimum vertical descent that is required in descending
from $P$ before ascending a higher, neighboring peak $P^\prime$, where $P^\prime$ can
lie to either side of $P$.
Fig.~\ref{masspeakplottop} illustrates
this concept for two different peaks.  In Fig.~\ref{zoomin} we illustrate how using prominence instead of
height to identify physical peaks can eliminate extraneous peaks that are artifacts of the detector's finite angular resolution.
The pictured jet has two distinct hard subjets separated by a single angular scale $\Delta R$.  Since one of the subjets
has its energy deposited in two neighboring calorimeter cells, the angular structure function $\Delta \mathcal{G}(R)$ exhibits
two distinct peaks in the neighborhood of $R=\Delta R$.  Only one of the two peaks has a large prominence, and
so using prominence to select peaks in $\Delta \mathcal{G}(R)$ ensures that only a single angular scale near $R=\Delta R$ 
is identified.  In the following we will identify a peak in $\Delta \mathcal{G}(R)$
by demanding that its prominence exceeds a minimum value $h_0$.

So far we have described how to define two different jet observables from prominent peaks in $\Delta \mathcal{G} (R)$.
The first is $n_p$, the number of prominent peaks in $\Delta \mathcal{G} (R)$.  The second is the various
angular scales $R_{i*}$ at which prominent peaks are located.  It remains
to define a jet observable that corresponds to cliff heights in $\mathcal{G} (R)$.   The magnitude of a cliff's height in
$\mathcal{G} (R)$ will map onto the height of the corresponding peak in $\Delta \mathcal{G} (R)$.  This
height is determined by the invariant mass of (typically) two hard subjets separated by an angular distance $R=R_{i*}$.  
For each prominent peak in $\Delta \mathcal{G} (R)$
with height $\Delta \mathcal{G} (R_{i*})$ we define the partial mass $m(R_{i*})\equiv m_{i*}$ as
\beq
m_{i*}^2\equiv \sqrt{\pi dR^2} \frac{ \Delta \mathcal{G} (R_*) \ \mathcal{G} (R_*)}{R_*} \ \mu_J^2
\eeq{partmass}
where we have used Eq.~\ref{masspeak} to extract the (appropriately normalized) numerator of the angular structure function.  Here
\beq
\mu_J^2 = \sum_{i \ne j}p_{T i}p_{T j}\Delta R_{ij}^2
\eeq{mu2}
is the denominator of $\mathcal{G}(R)$ in
Eq.~\ref{corrfunc} and is approximately equal to the squared jet mass $m_J^2$.  
To see the physics that is encoded in the partial mass consider a jet with two infinitely narrow, hard
subjets separated by an angular distance $\Delta R$ and with transverse momenta $p_{T1}$ and $p_{T2}$.  
This jet will exhibit a single prominent peak in $\Delta \mathcal{G} (R)$ at $R=\Delta R$.  The corresponding partial mass $m_*$ 
will be given by $m_*^2 =  p_{T1}p_{T2}\Delta R^2 \approx 2 p_1 \cdot p_2$.\footnote{Note that for two subjets $j_1$ and $j_2$ that
are not infinitely narrow, the gaussian kernel in Eq.~\ref{masspeak} introduces some amount of smearing in the partial mass.}  Thus
the partial mass is a measure of the mass at a particular angular scale.  For
a jet whose substructure is determined by a heavy particle decay, the partial masses will be fixed by the 
kinematic constraints of the decay.  This observation will be explored further in Sec.~3 in the context of top tagging.

Now that we have defined $n_p$, $R_{i*}$, and $m_{i*}$, we can ask how these jet observables
characterize the substructure of a jet.  First, for an idealized jet composed of $n_s$ hard, narrow subjets with each
pair of subjets separated by distinct angular scales $R_{i*}$, we expect the number of peaks $n_p$ 
to be given by
\beq
n_p=n_p^{\rm max} \equiv {n_s \choose 2}
\eeq{nsubjets}
In general this equality becomes an inequality $n_p\le n_p^{\rm max}$ for jets whose substructure is less clean.  For example,
if some of the $n_s$ subjets are wide or if some of the angular separations are approximately degenerate, then $\Delta \mathcal{G} (R)$
may exhibit fewer than $n_p^{\rm max}$ prominent peaks.  When a prominent peak is resolvable, however, the resulting angular
scale $R_{i*}$ corresponds to an angular separation between two or more hard substructures in the jet.  
For a QCD jet, the distribution of prominent peaks should be roughly uniform in $R$, since QCD is approximately
scale invariant.  For a jet that is initiated by a heavy particle decay, the angular scales $R_{i*}$ will be peaked at 
values characteristic of the decay kinematics of the heavy particle.  The corresponding partial masses will be correlated
to mass scales intrinsic to the heavy particle decay.  In contrast, for QCD jets the partial masses
will be peaked at small values, as determined by the soft and collinear singularities of QCD.

\begin{figure}[t]
\centering
\subfigure[]
{
    \includegraphics[width=7.3cm]{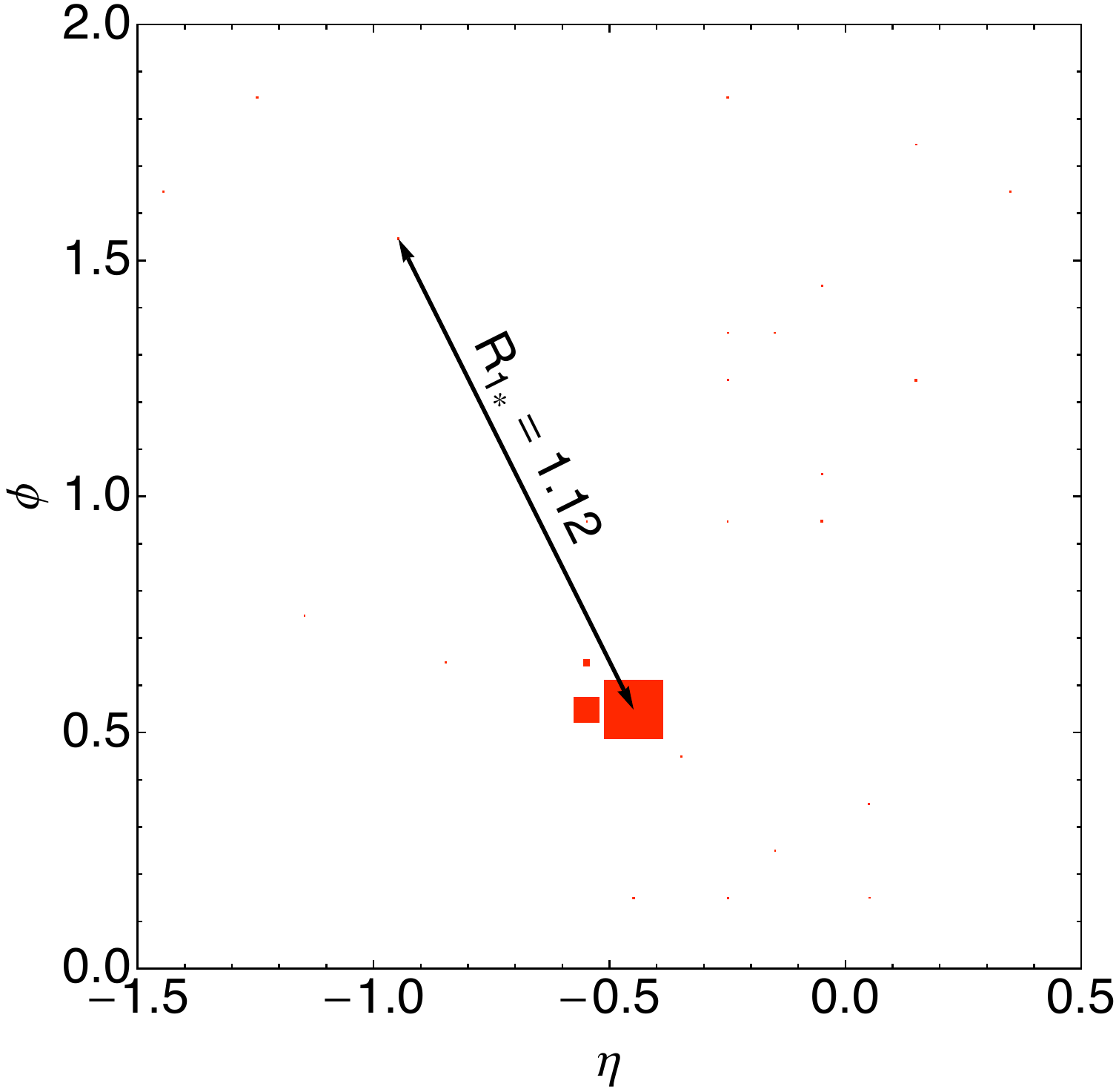}\label{legoqcd}
}
\hspace{.1cm}
\subfigure[]
{
    \includegraphics[width=7.0cm]{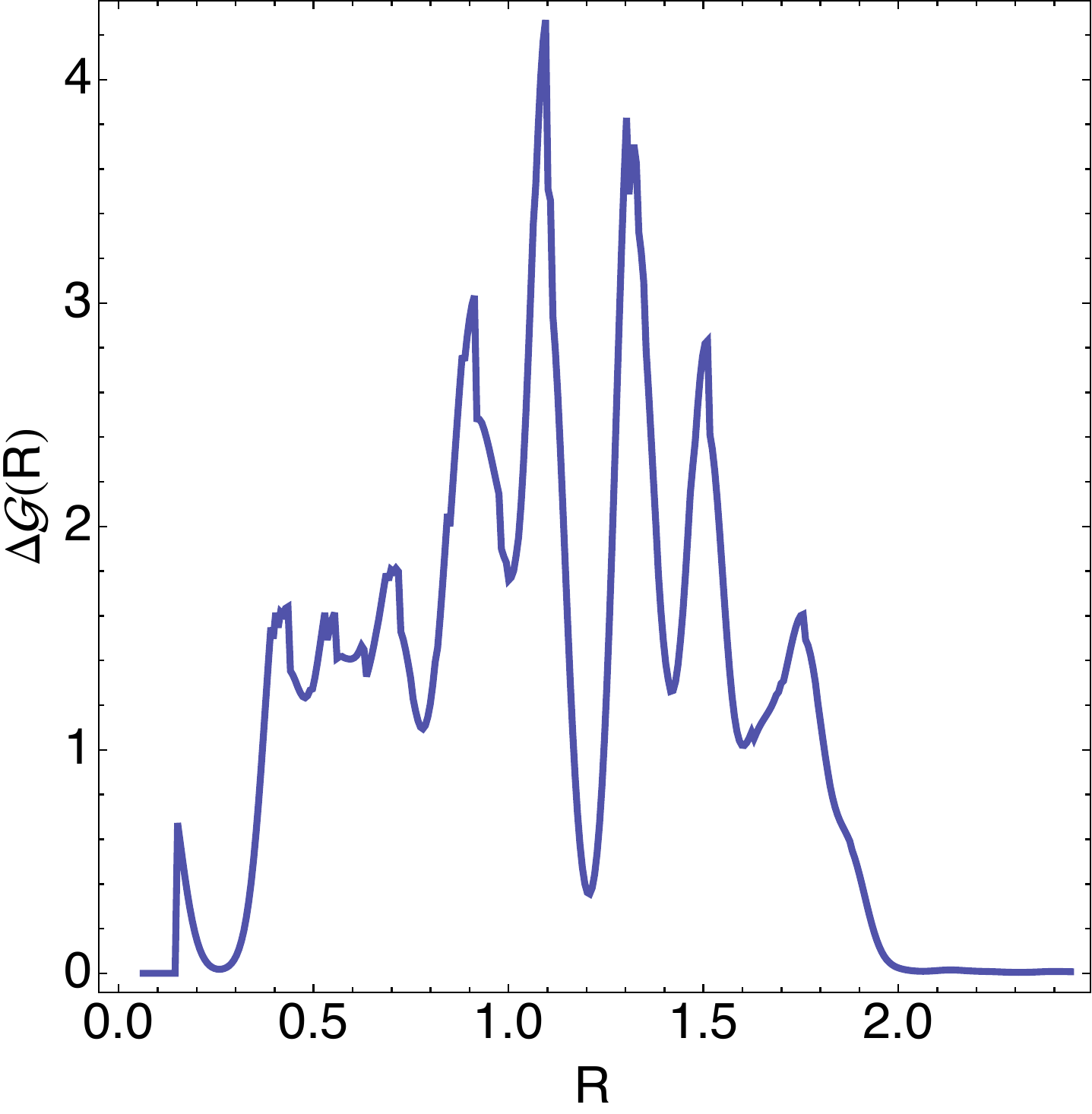}\label{masspeakplotqcd}
}
\caption{ {\bf (a)} $p_T$ plot and  {\bf (b)} angular structure function $\Delta \mathcal{G}(R)$ for a QCD jet with diffuse 
substructure and $p_T\sim600\ $GeV. 
In the $p_T$ plot, the small cell at the end of the arrow is so soft that it is barely visible. 
Prominent peaks in $\Delta \mathcal{G}(R)$ are distributed approximately uniformly in $R$.  For a minimum prominence of 4.0, $\Delta \mathcal{G} (R)$ has a single peak at $R_{1*}=1.09$.  Note 
the scale of $\Delta \mathcal{G}(R)$ as compared to the top jet in Fig.~\ref{masspeakplottop}.}
\label{qcdjet}
\end{figure}

Some of the foregoing discussion is illustrated in Figs.~\ref{topjet} and \ref{qcdjet}.
In Fig.~\ref{topjet} we show a boosted top jet with a clean three-pronged substructure.  In the $p_T$ plot in Fig.~\ref{legotop} the distances $R_{i*}$ 
between the three hardest cells are indicated.  From Fig.~\ref{masspeakplottop} we
see that it is these same three angular scales that show up as
prominent peaks in the angular structure function $\Delta \mathcal{G} (R)$.  Less prominent
peaks correspond to soft-hard correlations in the jet.   The substructure of the QCD jet in 
Fig.~\ref{legoqcd}
is quite different, with a single hard core surrounded by soft diffuse radiation.  The mass of
the jet is largely due to these soft, wide-angle emissions, and the most prominent
peak in $\Delta \mathcal{G} (R)$ corresponds to correlations between the
hard core of the jet and one such emission.  Prominent peaks in $\Delta \mathcal{G} (R)$
for this QCD jet are distributed approximately uniformly in $R$, as expected.  

The close correspondence between structure in the $p_T$ plots
apparent by eye and the structure identified by the angular structure function $\Delta \mathcal{G}(R)$ is encouraging.  To investigate
the effectiveness of this procedure more thoroughly will require testing it against a concrete application,
where the characteristics of the observables $n_p$, $R_{i*}$, and $m_{i*}$ can be explored in greater
detail.  A good testbed will involve jets with complex substructure.  For this
reason we choose to construct a top tagging algorithm as a first application.

\section{Top tagging}

\begin{figure}[t]
\centering
\subfigure[]{
    \includegraphics[width=7.2cm]{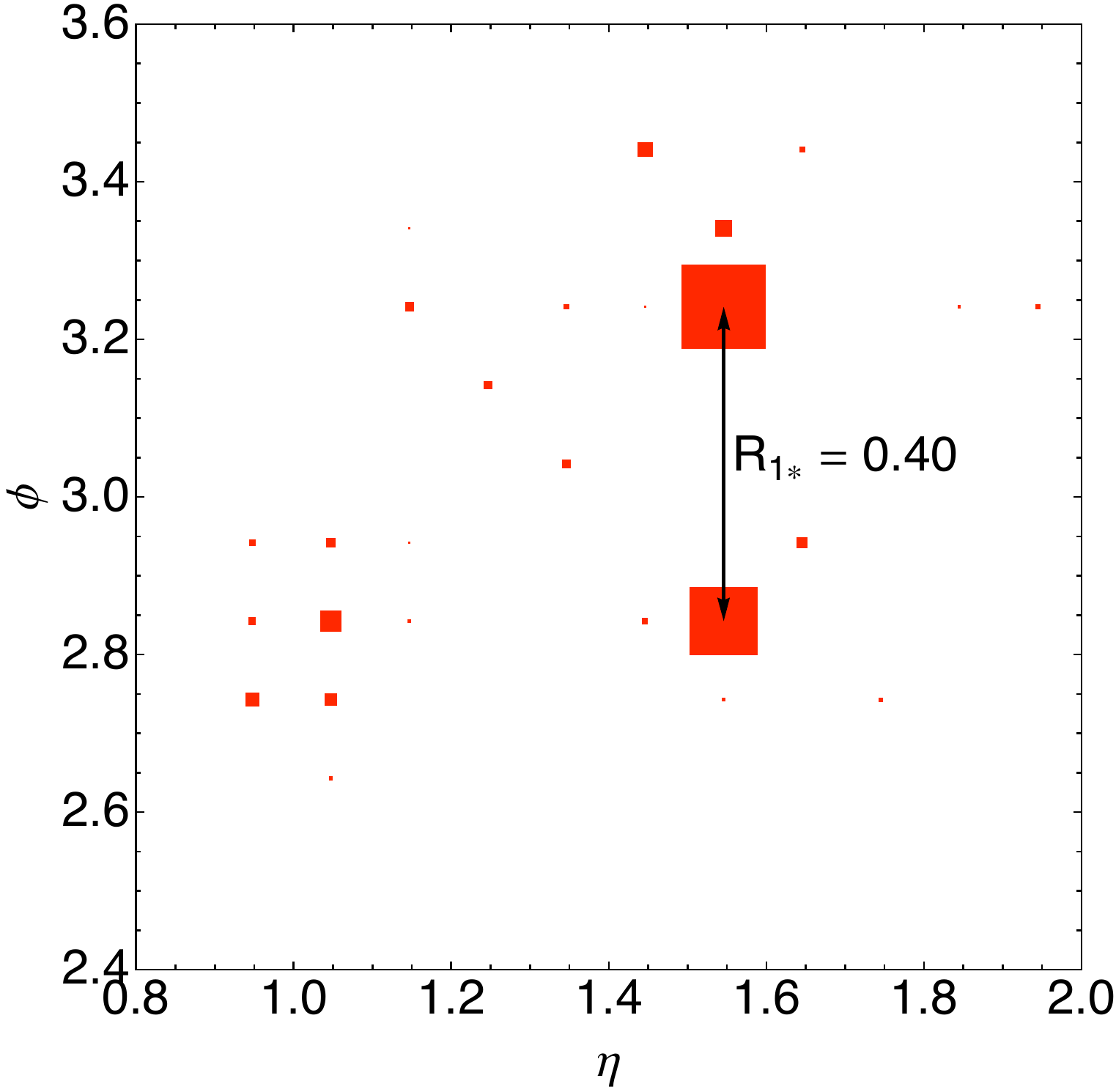}\label{legotop1}
}
\hspace{.1cm}
\subfigure[]
{
    \includegraphics[width=7.0cm]{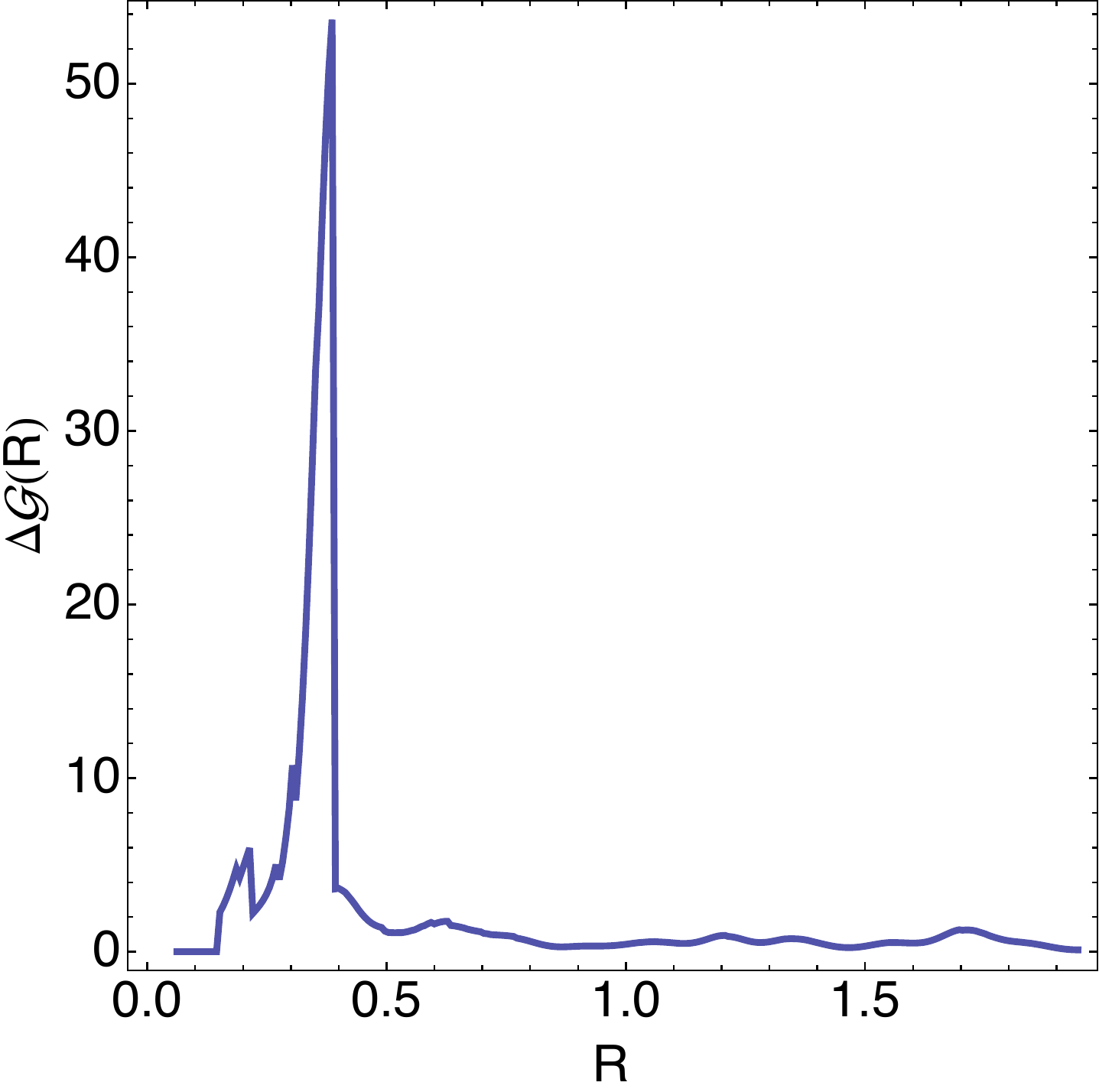}\label{masspeakplottop1}
}
\caption{{\bf (a)} $p_T$ plot and {\bf (b)} angular structure function $\Delta \mathcal{G}(R)$ for a top jet with $p_T \sim 500$ GeV.   The decay products of the $W^\pm$ are not individually resolved, with most of the radiation from the $W^\pm$ ($\phi \sim 2.8$) contained within a single, hard cell.  For a minimum prominence of 4.0, $\Delta \mathcal{G} (R)$ has a single peak at $R_{1*}=0.39$.}
\label{topjet2}
\end{figure}

If every top jet had the clean three-pronged structure apparent in Fig.~\ref{legotop} then
constructing an efficient top tagger would be straightforward.  In practice, reconstruction
of the top is complicated by a number of factors, including: (i) the finite resolution
of the detector, which degrades mass and angular resolution; (ii) collinear radiation,
which can make it difficult to resolve subjets initiated by hard partons that are close together;
and (iii) the boost from the top rest frame to the lab frame, 
which can result in decay products that are soft or overlap with one another.  As a consequence,
many top jets will have fewer than three prominent peaks in their angular structure functions.  
For example, in Fig.~\ref{topjet2} we show an example of a top jet in which the $W^\pm$
decay products do not exhibit a clean two-pronged structure. As a result $\Delta \mathcal{G} (R)$
only has a single prominent peak corresponding to mass correlations between the  $W^\pm$ and 
the $b$ subjet.  Constructing a tagger with high signal efficiencies will therefore require considering top jets with
fewer than three prominent peaks in their angular structure functions.  

This suggests that the following
procedure could result in an efficient top tagging algorithm.  
Fix a minimum prominence $h_0$.  For each candidate jet, calculate the angular structure
function and identify the number of peaks $n_p$ with prominences exceeding $h_0$.  
Reject candidate jets with $n_p=0$ or $n_p>3$ and sort the rest into bins with $n_p=1,2,3$.  Then apply separate sets of cuts 
to the $R_{i*}$ and $m_{i*}$ in each bin.
This procedure has the advantage that candidate jets are being sorted with respect to their observed
topologies.  For example, top jets in which the decay products of the $W^\pm$ are merged will be treated
differently from top jets that exhibit a clean three-pronged substructure.  In each bin cuts
will be applied to the observables available from the identified substructure, and the cuts
can be separately optimized to reflect the diversity of actual tops.  By not requiring candidate jets
to have the substructure of an idealized top jet with three distinct prongs, the top tagger can
be more accommodating towards ``ugly duckling'' tops and thus attain higher signal efficiencies\footnote{A similar
flexibility is found in the multi-body filtering employed by the {\tt HEPTopTagger} \citep{hep,stoprecon}.}.

The outline of this section is as follows.  In Sec.~\ref{secobs} we discuss distributions of the observables $R_{i*}$ and $m_{i*}$ for
top jets and QCD jets.  In Sec.~\ref{secalgo} we present the details of our top tagging algorithm.
In Sec.~\ref{secresults} we describe the Monte Carlo used to test the top tagger as well as the
performance of the algorithm.  

\subsection{Observables}
\label{secobs}

\begin{figure}
\centering
    \includegraphics[width=15.1cm]{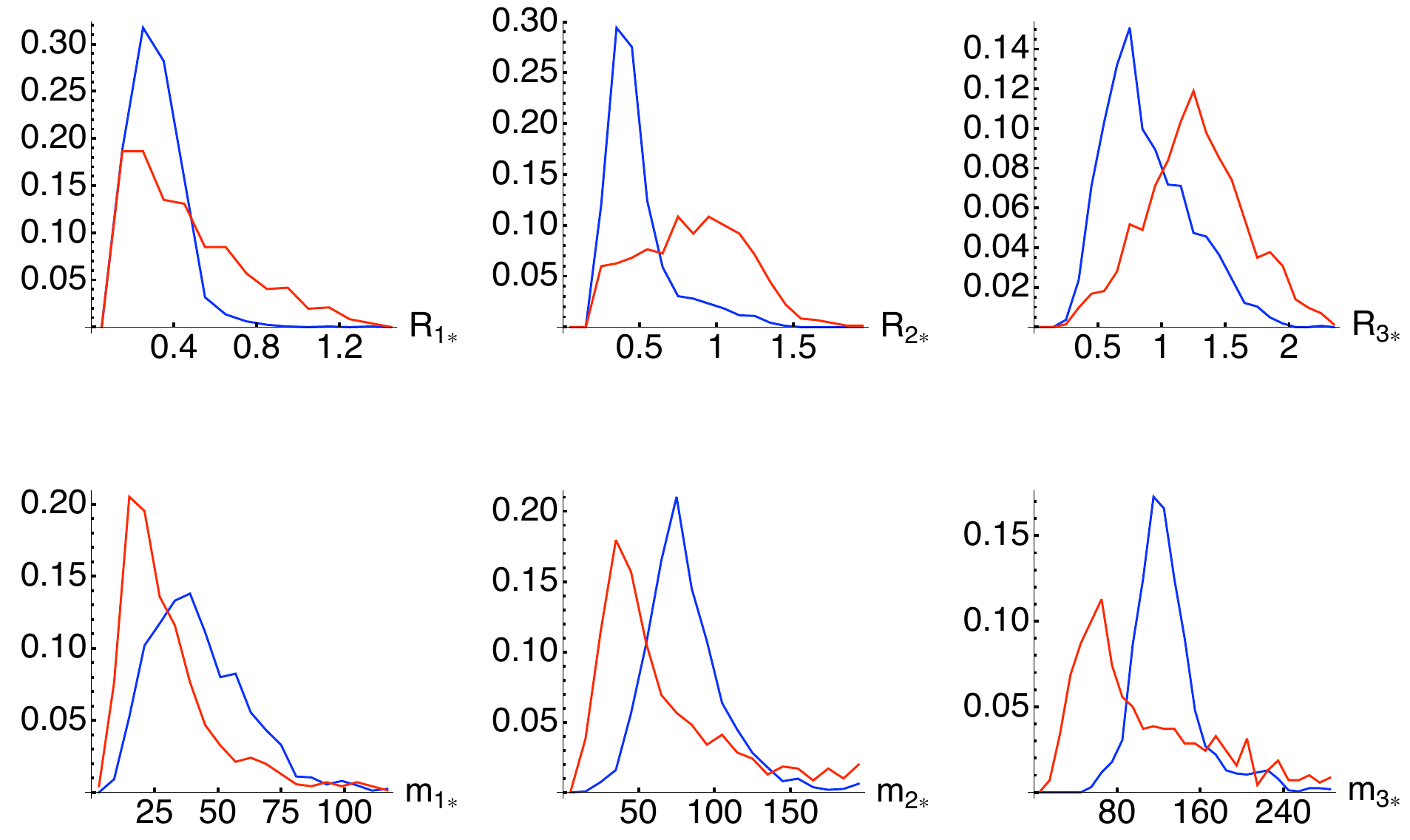}
\caption{Distributions for observables in the $n_p=3$ bin with 500 GeV $\le p_T \le $ 600 GeV.
 Distributions for top jets (QCD jets) are shown in blue (red).  Angular scales $R_{i*}$
  and partial masses $m_{i*}$ are ordered so that 
$R_{1*}\le R_{2*}\le R_{3*}$.  For QCD the $R_{i*}$ distributions are consistent 
with scale-invariant emission, while the $m_{i*}$ distributions peak towards small 
partial masses.  For tops the $R_{i*}$ and  $m_{i*}$ distributions are peaked at 
angular and mass scales characteristic of top decay kinematics.   Distributions are normalized
to unity.}\label{n3rm}
\end{figure}

\begin{figure}
\centering
    \includegraphics[width=8.8cm]{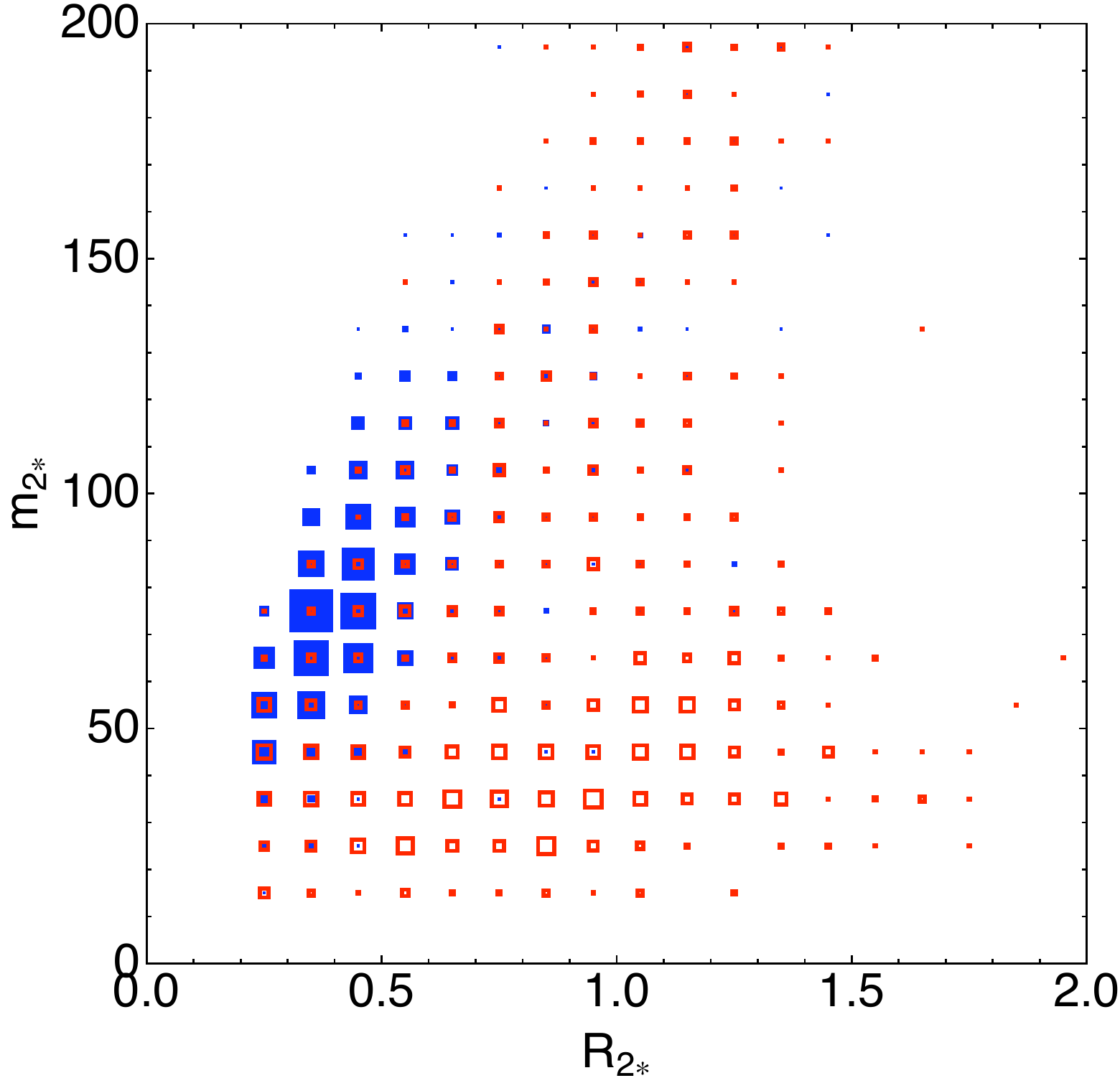}
\caption{Correlations between $R_{2*}$ and $m_{2*}$ in the $n_p=3$ bin with 500 GeV $\le p_T \le $ 600 GeV.  For the top kinematic constraints imply 
strong correlations between $R_{2*}$ and $m_{2*}$,
while for QCD jets the two are uncorrelated.
Correlations for top jets (QCD jets) are depicted in blue (red).}\label{r2m2corr}
\end{figure}

To set the stage for the top tagging algorithm defined in the next section, 
we first discuss what sort of top jet discrimination is available from the observables $R_{i*}$ and $m_{i*}$.
In Fig.~\ref{n3rm} we illustrate distributions for these observables 
in the $n_p=3$ bin.  
For top jets the
kinematic constraints of the top decay in conjunction with the boost to the lab frame account for the basic features
(see appendix A for details).  Identifying
the smallest $R_*$, i.e.~$R_{1*}$, with the angle between the $b$ subjet and the closer of the $W^\pm$ subjets, 
we expect that $R_{1*} \sim 0.25$ for this 500 GeV $\le p_T \le $ 600 GeV bin.  Similarly, identifying
$R_{2*}$ with the angle between the two $W^\pm$ subjets and $R_{3 *}$ with the angle between the 
$b$ subjet and the further of the $W^\pm$ subjets, we
expect that $R_{2*}\sim 0.50$ and $R_{3*}\sim 0.75$.   With these identifications for the three peaks, the predictions
for the partial masses become $m_{1*}\sim 50$ GeV,  $m_{2*}\sim m_W$, and $m_{3*}\sim140$ GeV.
These predictions for the $R_{i*}$ and $m_{i*}$ match up well with the distributions in Fig.~\ref{n3rm}, although in practice the
corresponding identifications only hold on the average.  Note that the kinematic constraints of the top quark decay imply strong
correlations between $R_{i*}$ and $m_{i*}$ for each $i$.  This is illustrated in Fig.~\ref{r2m2corr}, where $R_{2*}$ has been plotted
against $m_{2*}$ in the $n_p=3$ bin.  For QCD jets $R_{2*}$ and $m_{2*}$ are uncorrelated.

In contrast to top jets, QCD jets have no intrinsic scales.  
Since QCD is approximately scale invariant and the derivative in $\Delta \mathcal{G} (R)$ is with respect to $\log R$,
we expect the $R_{*}$ distributions to be approximately uniform.  Imposing
the ordering $R_{1*}\le R _{2*}\le R_{3*}$  then has the consequence that the $R_{1*}$ distribution should peak at $R=0$, 
the $R_{2*}$ distribution should peak at intermediate $R$, and the $R_{3*}$ distribution should peak towards large $R$. 
This is consistent with what is seen in Fig.~\ref{n3rm}, up to edge effects at large $R$ in the $R_{3*}$ distribution.  The partial masses of QCD jets
are peaked towards small $m_{i*}$, as we expect given that the physics of $m_{i*}$ is qualitatively similar to the physics
of jet masses $m_J$. 

\begin{figure}
\centering
    \includegraphics[width=9.1cm]{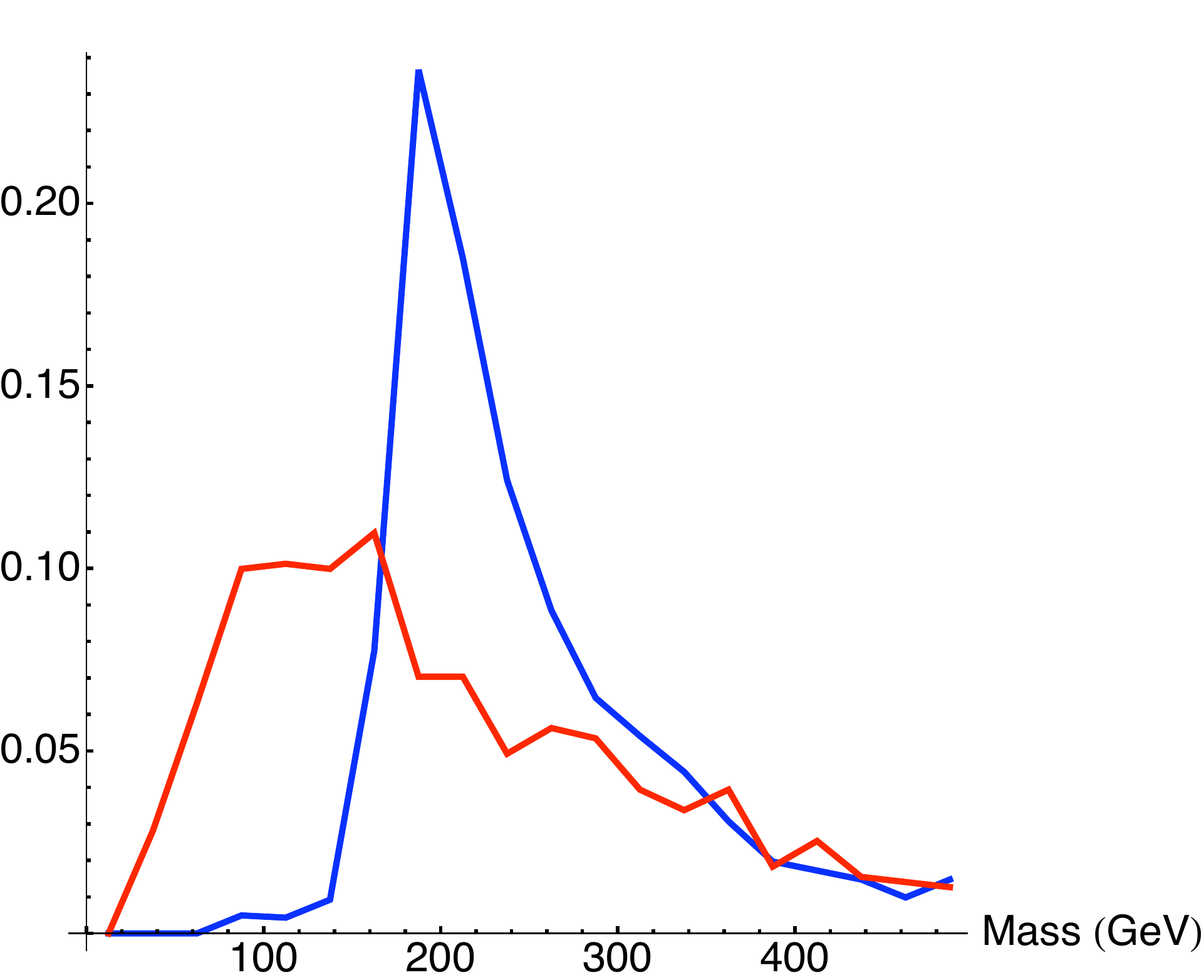}
\caption{The jet mass $m_J$ for tops (blue) and QCD (red) in the $n_p=3$ bin with 500 GeV 
$\le p_T \le $ 600 GeV.}\label{massdist3}
\end{figure}

\begin{figure}
\centering
    \includegraphics[width=15.1cm]{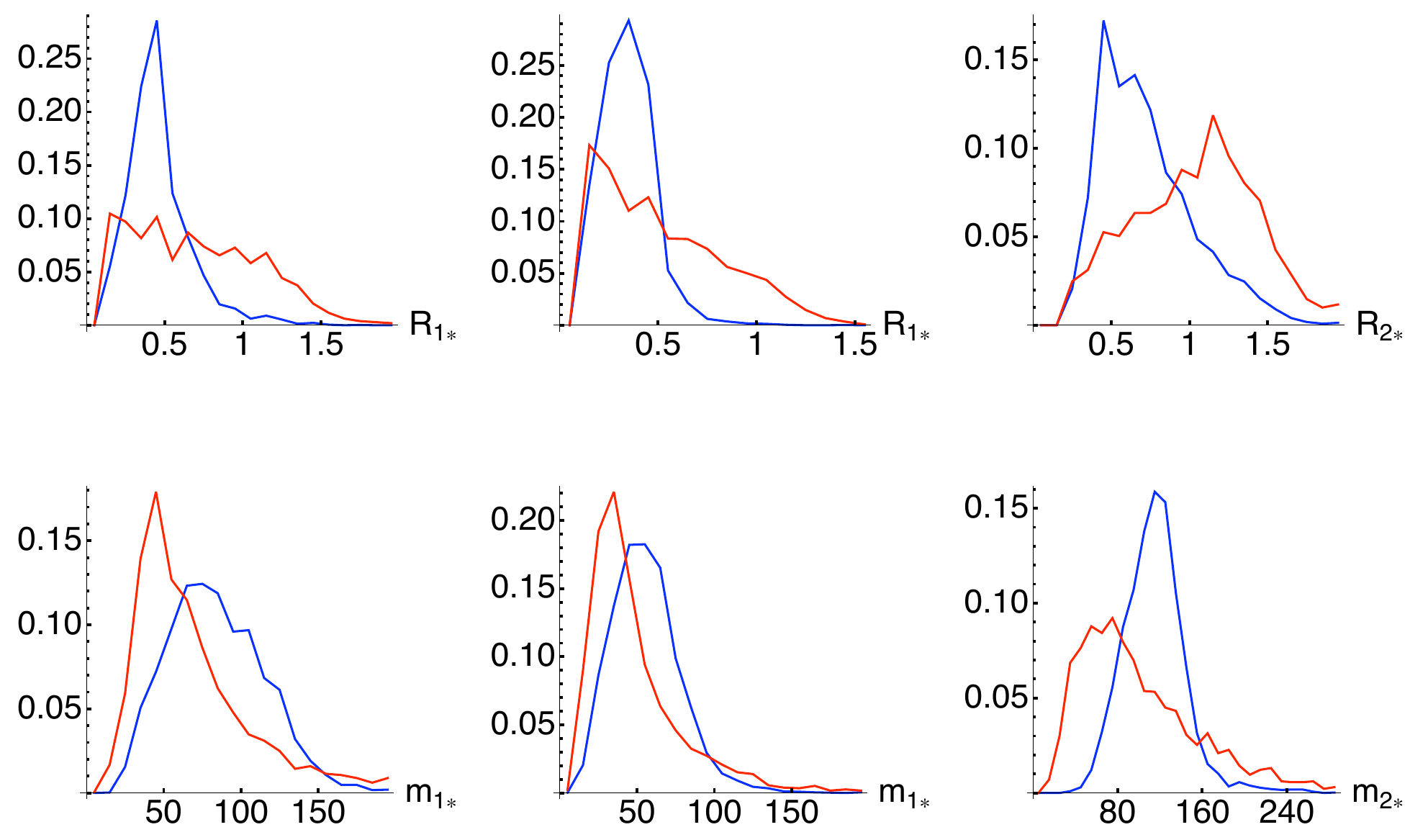}
\caption{Distributions for $R_{i*}$ and $m_{i*}$ in the $n_p=1,2$ bins with 500 GeV $\le p_T \le $ 
600 GeV.  
The leftmost column is $n_p=1$, and the two rightmost columns are $n_p=2$.}\label{n12plots}
\end{figure}

The features of the
distributions in the $n_p=1,2$ bins are qualitatively similar, see Fig.~\ref{n12plots}.  Here it is less clear what
identifications to make for the different peaks, and it is likely that there is a fair amount of mixing between
different decay topologies.     
In any case the observables derived from $\Delta \mathcal{G} (R)$ in the $n_p=1,2$ bins
make effective discriminants between top jets and QCD jets, although more discrimination is available in the $n_p=3$ bin.
The distributions for $R_{1*}$ and $m_{1*}$ in the $n_p=1$ bin are consistent with correlations between the $W^\pm$ subjets $j_{W1}$ and $j_{W2}$; one possibility is that for these
top jets the $b$ subjet is too soft to yield prominent peaks.  
The distributions for the $n_p=2$ bin are consistent with correlations
between the $b$ subjet and each of the two $W^\pm$ subjets; one possibility is that for these top
jets the $W^\pm$ subjets $j_{W1}$ and $j_{W2}$ are nearly merged so that correlations between $j_{W1}$ and $j_{W2}$ 
do not result in any prominent peaks.  

\begin{figure}[t]
\centering
    \includegraphics[width=8.1cm]{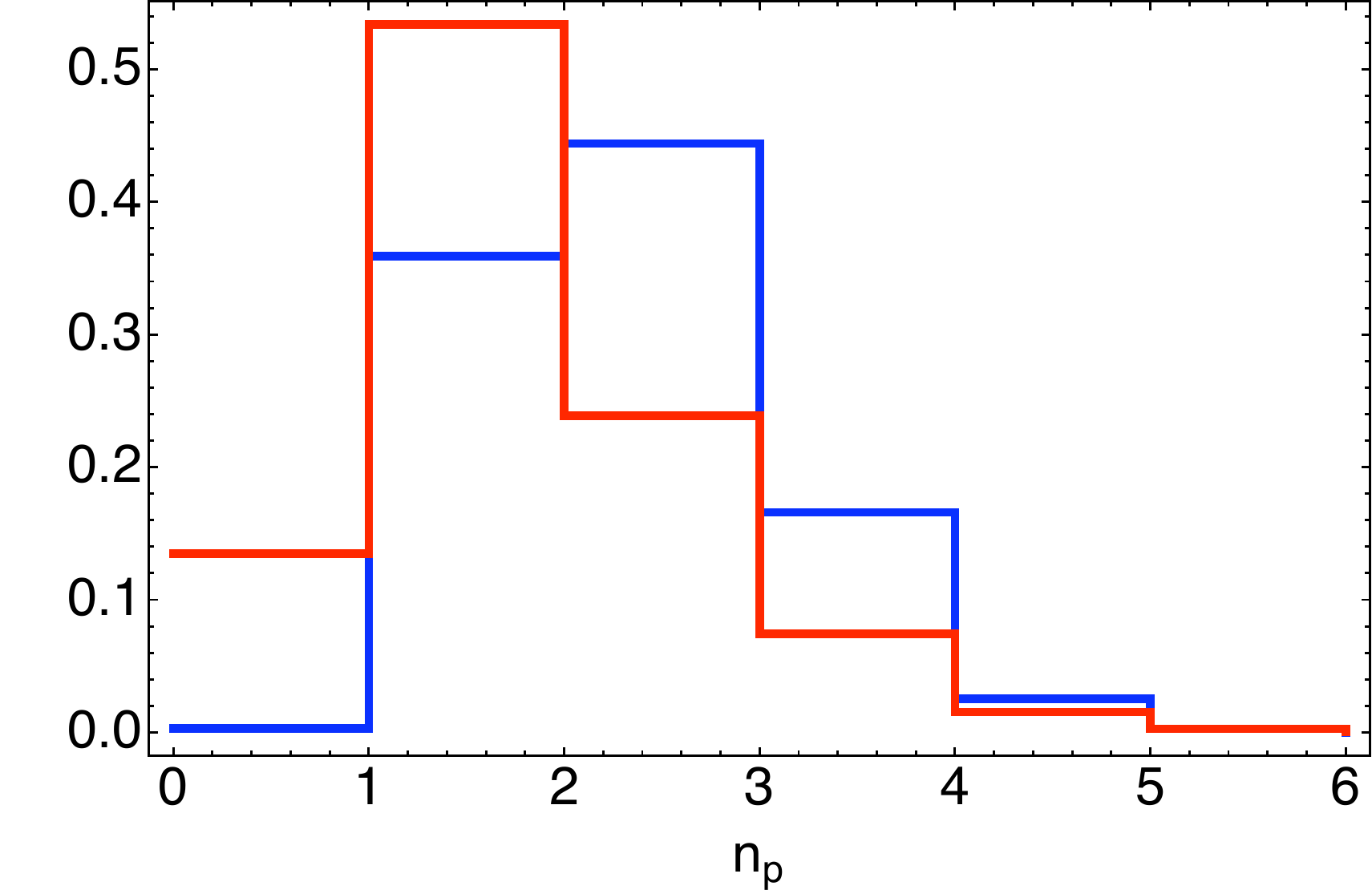}
\caption{Fractions of top jets (blue) and QCD jets (red) that have $n_p$ prominent peaks.  
Here the minimum prominence is $h_0=4.0$ and 500 GeV $\le p_T \le $ 600 GeV.  These fractions exhibit only a small dependence on $p_T$.}
\label{peaknoplot}
\end{figure}

\subsection{An algorithm}
\label{secalgo}

The distributions in Figs.~\ref{n3rm}-\ref{n12plots} suggest
that imposing cuts on $m_J$, $R_{i*}$, and $m_{i*}$ could lead to effective discrimination between top jets and QCD jets.
To test this we employ the following top tagging algorithm.  Using the CA algorithm, cluster the event into fat jets with $R=1.5$.
Although a more advanced version of the tagger could benefit from using variable $R$ (or a filtered jet mass $m_{\rm{filt}}$),
we leave the value of $R$ fixed for simplicity.
Before applying any cuts, first presort the candidate jets into $p_T$ bins of width $100$ GeV.  
Then for each candidate jet calculate $\Delta \mathcal{G} (R)$
and identify the number of peaks $n_p$ whose prominence exceeds a fixed minimum prominence $h_0=4.0$.  This
value of $h_0$ has been selected by scanning over a range $h_0 \in [1.0, 10.0]$ and choosing $h_0$ to minimize
the background efficiency over a wide range of $p_T$ and signal efficiencies. Within
each $p_T$ bin further sort the candidate jets into three peak bins ($n_p =1,2,3$), throwing out jets with $n_p=0$ or
$n_p>3$.  This $n_p$ cut removes a sizable fraction ($\sim 15\%$) of QCD jets, while rejecting only $\sim 3\%$ of top
jets, see Fig.~\ref{peaknoplot}.
For discrimination between top jets and QCD jets to be most effective one would like to disentangle the correlations between the observables 
as much possible; for simplicity, however, we choose to make rectangular cuts in the space of observables.  In particular, in the $n_p=3$ bin we
choose to impose cuts on six of the seven available observables, excluding $m_{1*}$, which is the least discriminating observable.  
More specifically, we impose the following cuts:
\begin{enumerate}
\item $m_J > m_{\rm{t\;min}}$
\item $R_{1*} < R_{1*}^{\rm{max}}$, \
 $R_{2*} < R_{2*}^{\rm{max}}$, \
 $R_{3*} < R_{3*}^{\rm{max}}$
\item $m_{2*} > m_{2*}^{\rm{min}}$, \
 $m_{3*} > m_{3*}^{\rm{min}}$
\end{enumerate}
A candidate jet that passes this set of cuts is tagged as a top jet.  
In the $n_p=1, 2$ bins we employ the corresponding set of cuts, except in contrast to the $n_p=3$ bin, we make use of all of the observables.
Also, we impose an additional cut $m_J < m_{\rm{t\;max}}$ in the $n_p=1$ bin only, since the smaller number of observables in the 
$n_p=1$ bin (three) means that imposing this cut does not substantially increase the computer time needed to find optimal cuts.  For the moment we leave the values of the cuts unspecified; this will be addressed in the next section.

\subsection{Results}
\label{secresults}

We use two different event samples for evaluating the performance of the top tagger.  
These event samples (from $pp$ collisions with center of mass energy of 7 TeV) belong to a set of benchmark event samples that have been made publicly available by participants of the BOOST 2010 workshop \citep{BOOSTEvents}.  
The first event sample is generated by {\tt HERWIG}~6.510 \citep{herwig} with the underlying event simulated by {\tt JIMMY} \citep{jimmy}, which has been configured with a tune used by ATLAS.  
The second is generated by {\tt PYTHIA}~6.4 \citep{pythia} with $Q^2$-ordering and the `DW' tune for the underlying event.
See \citep{BOOST2010} for
more details.  Unless noted otherwise, all results presented in this paper make use of the {\tt HERWIG} event samples; 
the {\tt PYTHIA} event samples were used as crosschecks.  For signal jets we use the hardest jet in
 each event of a Standard Model hadronic $t\bar t$ sample, excluding
jets with $|\eta| > 2.5$.  For background jets we use the hardest jet in each event of a Standard 
Model dijet sample, 
again excluding jets with $|\eta| > 2.5$.  For both event samples there are ${\cal O}(10^4)$ events in each $p_T$ bin
of width $100$ GeV.
For jet clustering we use the CA algorithm \citep{CAalg} with $R=1.5$ as implemented by  
{\tt FastJet~2.4.2} \citep{fastjet}.  
In order to simulate the finite resolution of the ATLAS or CMS calorimeters, particles in each event 
are clustered into $0.1 \times 0.1$ cells
in $(\eta,\phi)$ and then combined into massless four-vector pseudoparticles that are fed 
into {\tt FastJet}.    
For each $p_T$ window the cuts are chosen to yield 
the smallest background efficiency $\epsilon_B$ at each fixed signal efficiency  $\epsilon_S$.  
This optimization is performed by a custom Monte Carlo code that finely
 samples the space of cuts.  Some sample values for the different cuts are given in Table \ref{cuttable}.

\begin{figure}
\centering
\subfigure[]
{    \includegraphics[width=7.00cm]{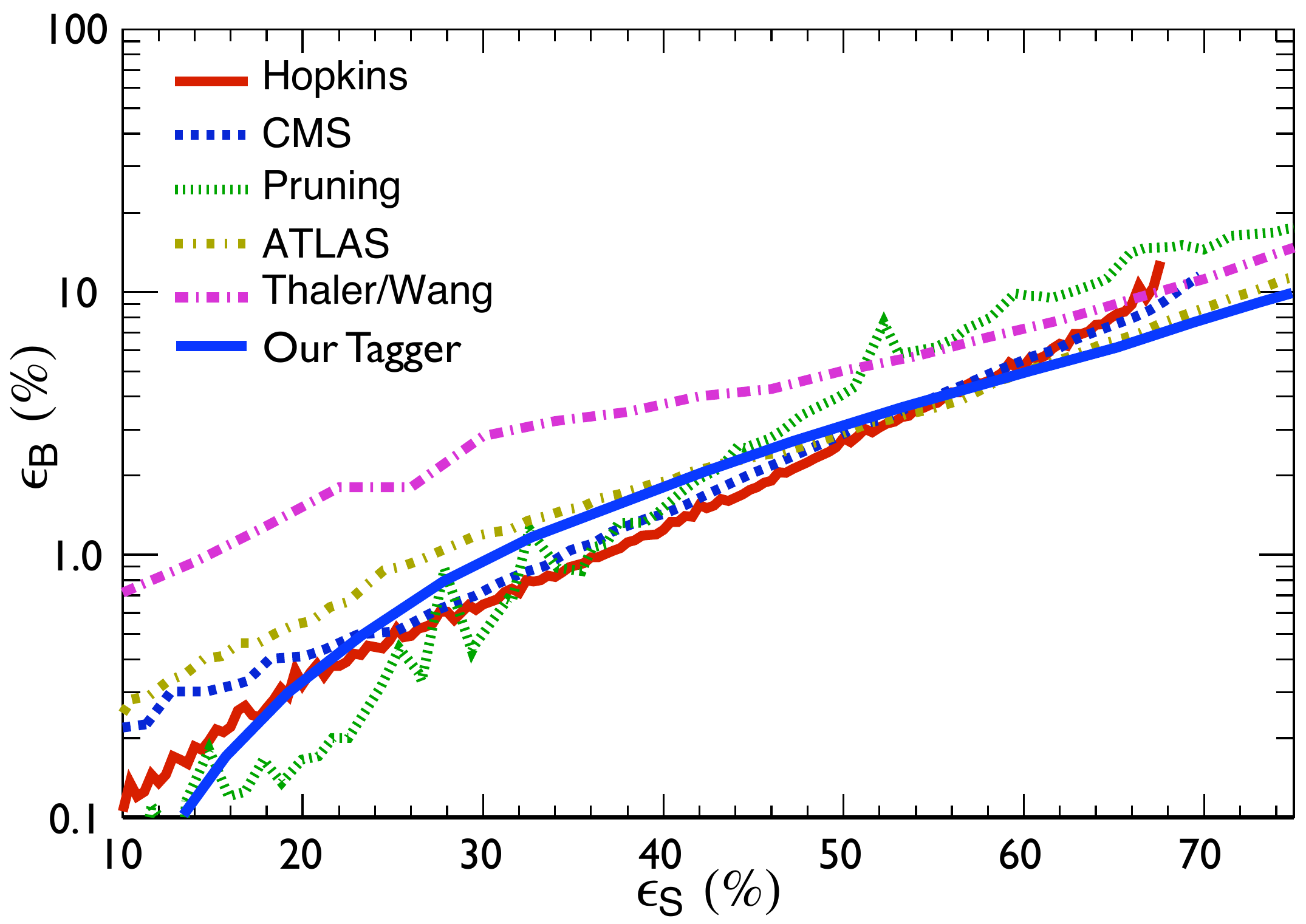}\label{svb500}}
\hspace{.1cm}
\subfigure[]
{    \includegraphics[width=7.60cm]{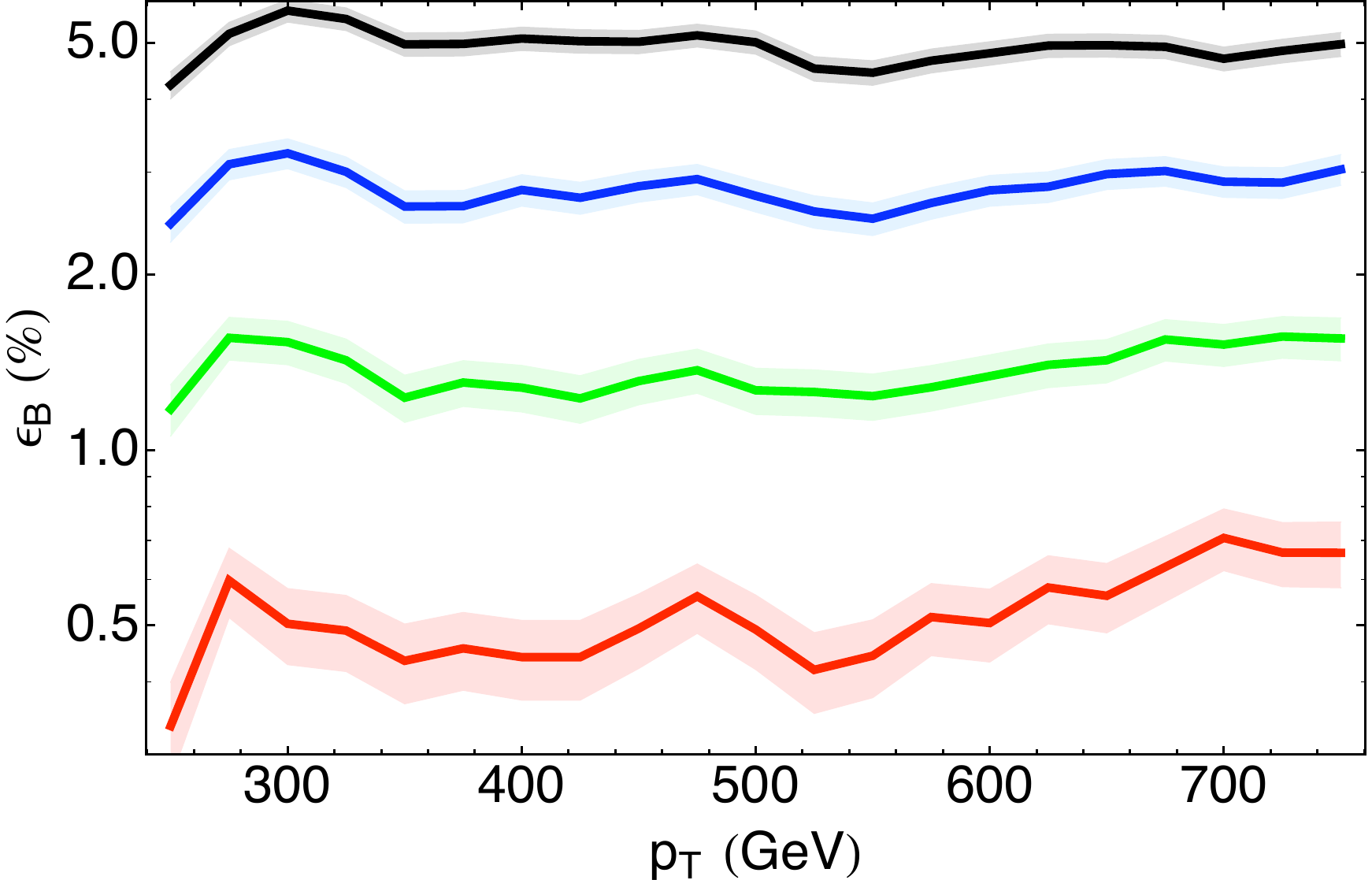}\label{sig50}}
\caption{The performance of the top tagger as given by the {\tt HERWIG} 
event samples.  The background efficiency vs.~signal efficiency for our
top tagger is compared to other algorithms in the literature in {\bf (a)}.  This figure is reproduced from 
\citep{BOOST2010} with the results from our tagger added.  Here the candidate jets have 
transverse momenta 500 GeV $\le p_T \le $ 600 GeV.  For Fig.~{\bf (a)} only, candidate jets
have been clustered with the anti-kT algorithm with $R=1.0$, as was done in the BOOST study.  As a consequence
the performance in {\bf (a)} is better than in {\bf (b)}, where the large jet radius degrades top mass resolution.  In {\bf (b)} the background efficiency is plotted
as a function of $p_T$ for signal efficiencies of $\epsilon_S=50\%$ (black), $40\%$ (blue), $30\%$ 
(green) and $20\%$ (red).  Efficiencies at a given $p_{T0}$ are calculated from a $p_T$ window of 
100 GeV centered at $p_{T0}$.  Note that, as a consequence, each point is not statistically 
independent.  Error bands are statistical.}
\end{figure}

\begin{table}
\begin{center}
\begin{tabular}{c | c c c c | c c}
$n_p=1$ & $m_{\rm{t\;min}}$ & $m_{\rm{t\;max}}$ & $R_{1*}^{\max}$ & $m_{1*}^{\min}$  & $\epsilon_S$(\%) & $\epsilon_B$(\%) \\
\hline 
$300-400$ GeV & $177$ GeV & $300$ GeV  & $0.96$ & $78$ GeV & 23.8 & 1.9 \\
$500-600$ GeV & $175$ GeV & $300$ GeV & $0.57$ & $74$ GeV & 27.0 & 2.6 \\
\end{tabular}
%\end{center}
\\[6pt]
%\begin{center}
\begin{tabular}{c | c c c c c | c c }
$n_p=2$ & $m_{\rm{t\;min}}$ & $R_{1*}^{\max}$ & $R_{2*}^{\max}$ & $m_{1*}^{\min}$ & $m_{2*}^{\min}$ &
 $\epsilon_S$(\%) & $\epsilon_B$(\%) \\
\hline 
$300-400$ GeV & $157$ GeV & $0.85$ & $1.59$ & $30$ GeV & $77$ GeV & 57.2 & 11.4 \\
$500-600$ GeV & $159$ GeV & $0.57$ & $1.00$ & $36$ GeV & $55$ GeV & 59.6 & 9.8 \\
\end{tabular}
%\end{center}
\\[6pt]
%\begin{center}
\begin{tabular}{c | c c c c c c | c c }
$n_p=3$ & $m_{\rm{t\;min}}$ & $R_{1*}^{\max}$ & $R_{2*}^{\max}$ & $R_{3*}^{\max}$ & $m_{2*}^{\min}$ & $m_{3*}^{\min}$ 
& $\epsilon_S$(\%) & $\epsilon_B$(\%) \\
\hline 
$300-400$ GeV & $102$ GeV & $0.81$ & $1.03$ & $2.11$ & $26$ GeV & $79$ GeV & 82.9 & 15.9 \\
$500-600$ GeV & $155$ GeV & $0.62$ & $0.66$  & $1.35$ & $46$ GeV & $73$ GeV & 73.6 & 7.9 \\
\end{tabular}
\end{center}
\caption{Sample optimized cut parameters at a (total) signal efficiency of $\epsilon_S=50\%$ for two
different $p_T$ bins.  In the rightmost column we
show the signal and background efficiencies obtained within each $n_p$ bin taken separately; i.e.~these
numbers do not take into account what fraction of candidate jets end up in each $n_p$ bin.  Signal efficiency
 increases substantially with $n_p$.}\label{cuttable}
\end{table}

\begin{figure}[t]
\centering
    \includegraphics[width=8.7cm]{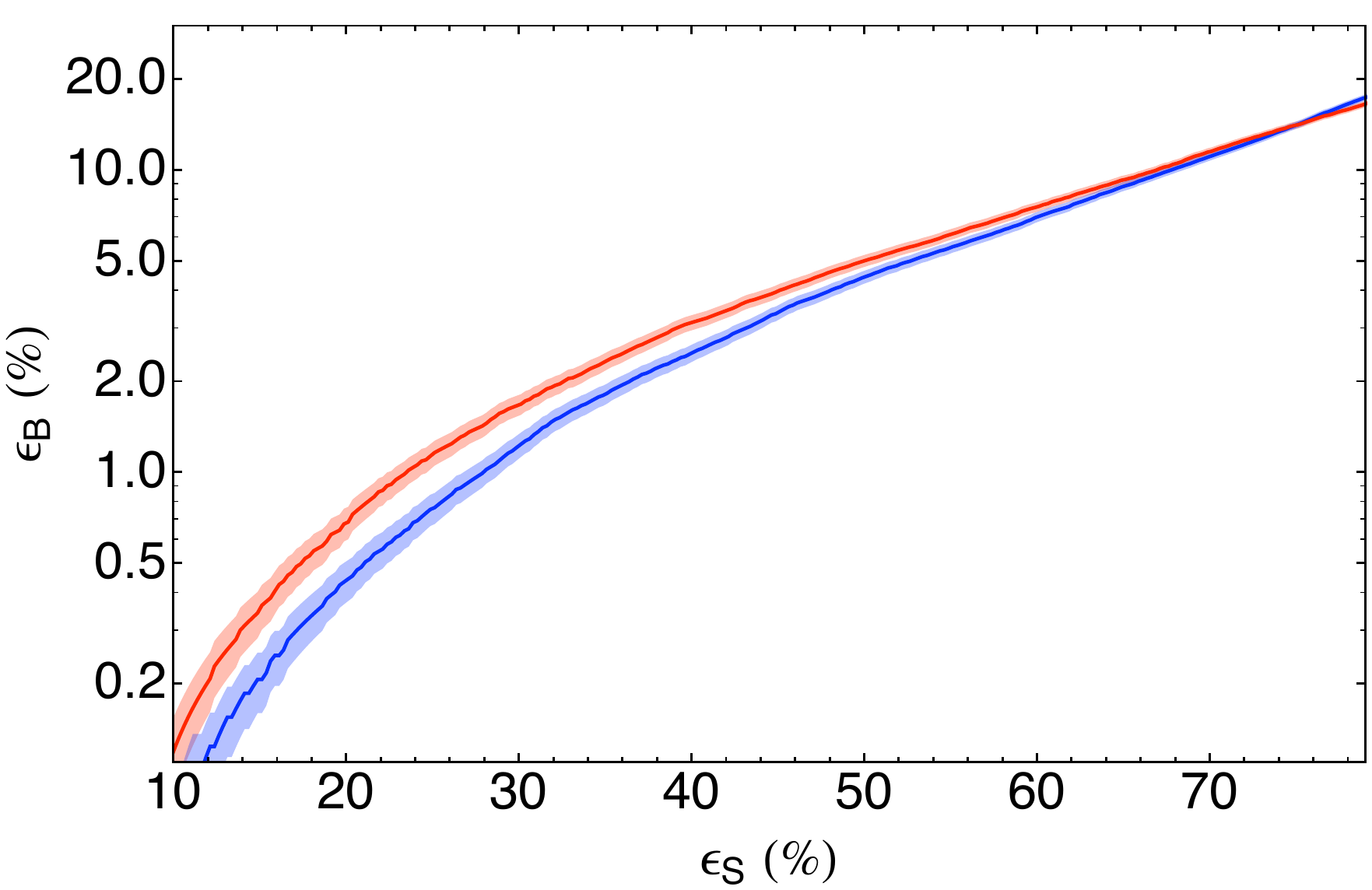}
\caption{Signal versus background efficiency curves for {\tt HERWIG} (blue) and {\tt PYTHIA}
(red) event samples in the 500 GeV $\le p_T \le $ 600 GeV $p_T$ bin.  Error bands are 
statistical.}
\label{pyhertest}
\end{figure}

In Fig.~\ref{svb500} and Fig.~\ref{sig50} we illustrate the performance of the top tagger.  
The performance is comparable to other top taggers in the literature \citep{toptag,cms1,cms2,cms3,EllisWalsh,brooijmans,atlas1,atlas2,thalerwang}, 
with $\epsilon_B \sim 5\%$ for 
$\epsilon_S =50\%$ and  $\epsilon_B \sim 0.5\%$ for $\epsilon_S =20\%$ \citep{BOOST2010}. 
For a fixed signal efficiency, the
background efficiency is approximately flat across the $p_T$ range we have 
tested, 200 GeV $\le p_T \le $ 800 GeV. 
In Table \ref{cuttable} we
see that in the $n_p=2$ and especially $n_p=3$ bins, where correspondingly more observables
are available for discrimination, the top tagger is able to attain large signal
 efficiencies.  Because the net
signal and background efficiencies are obtained by combining all three
 $n_p$ bins, the largest
contribution to $\epsilon_S$ is actually from the $n_p=2$ bin, since the
 plurality of top jets land in the
$n_p=2$ bin for $h_0=4.0$ (see Fig.~\ref{peaknoplot}).  For example, 
at $\epsilon_S=50\%$ and for
500 GeV $\le p_T \le $ 600 GeV about $55\%$ of tagged top jets come from the $n_p=2$ bin,
while about $20\%$ and $25\%$ come from the $n_p=1$ and $n_p=3$ bins, respectively.  Similarly,
the background efficiency is lowest in the $n_p=1$ bin; only QCD jets with two or three prominent
peaks do a good job of faking the substructure of a top jet.  For example, 
at $\epsilon_S=50\%$ and for
500 GeV $\le p_T \le $ 600 GeV about $32\%$, $54\%$, and $14\%$ of tagged QCD jets come from the $n_p=1$,
$n_p=2$, and $n_p=3$ bins, respectively, even though only about $31\%$ of QCD jets fall in the $n_p=$ 2 or 3 bins.

As a crosscheck in Fig.~\ref{pyhertest} we compare the performance of the top tagging algorithm between
the {\tt HERWIG} and {\tt PYTHIA} event samples.  We see that the background efficiency is
generally lower for {\tt HERWIG} than it is for {\tt PYTHIA}.  One possible reason for this is that
that although the cut parameters have been separately optimized for both event generators,
the parameters $h_0=4.0$ and $dR=0.06$ were optimized on the basis of the {\tt HERWIG}
event samples.  The {\tt HERWIG} and {\tt PYTHIA} event samples already disagree at the level
of the $n_p$ distributions, and this disagreement persists in the absence of the underlying event. 
This means that the typical prominence of peaks in $\Delta \mathcal{G(R)}$ differs between the
two event samples.  It would
be interesting to understand in detail which features of the two event
generators (the parton shower description, the underlying event model, etc.)~contribute to this disagreement.  Going
further in this direction, however, lies outside the scope of this paper. 
 
Given the large number of cut parameters that enter into the top tagging algorithm, overtraining is
a concern.  By training the cut parameters on a subset $A$ of the event samples and testing
the resulting cuts on subsets $B_i$ disjoint from $A$, we can get some idea for how susceptible 
the quoted efficiencies are to overtraining.  We find that the variation in the background efficiency 
$\epsilon_B$ (at fixed $\epsilon_S$) that results from this validation procedure is comparable to the 
quoted statistical uncertainties.  This additional uncertainty should be kept in mind when considering
the absolute performance of the top tagger.  Since precise estimates for background efficiencies
are made difficult by other uncertainties, such as those which enter the modeling of QCD backgrounds or 
detector mock-up, we do not consider overtraining any further.

Our simple mock calorimeter does not account for a variety of detector effects.  Recent
studies at the LHC (see e.g.~\citep{cms4}) suggest that Monte Carlo tools provide a fair description of the
performance of jet substructure algorithms.  Since the algorithm discussed in this paper relies on
kinematic observables, we suspect that the performance of the algorithm will not be
exceedingly sensitive to detector effects.  As a consequence the tagging efficiencies quoted above should
be fair estimates of what can be expected in the absence of pile-up.
Sensitivity to pile-up requires further study, and a full detector simulation
would be required to better understand the expected performance of the top tagging algorithm.  Aspects
of the algorithm may also be amenable to a sideband analysis.  In particular, by looking at regions of the 
$R_{i*}$-$m_{i*}$ plane (see Fig.~\ref{r2m2corr}) away from the signal region the shape of the background 
distributions can be extrapolated into the signal region.   

\section{Discussion}

By sorting jets according to the number of prominent peaks identified in their
angular structure functions $\Delta \mathcal{G}(R)$ and making rectangular cuts on the angular and
mass scales $R_{i*}$ and $m_{i*}$,
we have been able to construct an efficient top tagging algorithm.  Since the focus of this paper
has been to demonstrate that $\Delta \mathcal{G}(R)$ can be used to identify angular
and mass scales in jets, the particular algorithm we have described was chosen for its simplicity.  A number
of possible improvements to the algorithm suggest themselves, however, even leaving aside modifications
that are unrelated to the use of $\Delta \mathcal{G}(R)$.  One possible concern
is the large number of cut parameters that result from using three peak bins.  Given the strong correlations
between the $R_{i*}$ and $m_{i*}$ (see Fig.~\ref{r2m2corr}), one way to reduce the total number
of free parameters would be to consolidate some of the variables.  For example, one could
replace separate cuts on $R_{i*}$ and $m_{i*}$ with a single cut on $m_{i*}/R_{i*}$.  One could also
investigate different schemes for binning identified peaks in $\Delta \mathcal{G}(R)$.  For example,
the expected substructure of a top might be better captured by sorting into bins $\{n_{p0},n_{p1}\}$,
where bin $\{n_{p0},n_{p1}\}$ contains $n_{p0}$ peaks with prominence $P \ge h_0$ and $n_{p1}$ peaks with prominence $ h_1 \le P < h_0$.
The definition of the partial mass in Eq.~\ref{partmass}, which is most accurate
for narrow subjets, could be improved to better capture the invariant mass of wide subjets. 
The particular way in which we organize the observables $R_{i*}$ and $m_{i*}$ according to their ordering in $R$ as well
as the use of topographic prominence to identify peaks could
also be revisited.  Since $\Delta \mathcal{G}(R)$ defines a continuous number of observables, this list of possible
modifications could go on indefinitely, and it is interesting to ask whether our simple procedure
makes efficient use of the information available from $\mathcal{G}(R)$.  Going further in this direction, however,
lies outside the scope of this paper.

Although we have explored the use of the angular correlation function $\mathcal{G}(R)$ and the
angular structure function $\Delta \mathcal{G}(R)$
for the particular application of top tagging, the generality of the resulting procedure
suggests that it could be useful in a variety of different contexts.  It seems likely that procedures that make use of $\Delta \mathcal{G}(R)$
will be most effective when accurate reconstruction of angular scales is valuable.   
Some interesting possibilities include:
\begin{itemize}
\item using observables defined from $\Delta \mathcal{G}(R)$ to probe QCD; for example, measurements of $R_*$ or $n_p$
distributions for QCD jets could be compared against Monte Carlo calculations 
\item using $R_*$ distributions to search for new physics (angular bumps instead of mass bumps); this is attractive,
since accurate mass reconstruction is difficult
\item calculating $\Delta \mathcal{G}(R)$ for the event as a whole and using the identified angular scales to determine
an appropriate jet radius parameter $R$ event-by-event 
\item using $\Delta \mathcal{G}(R)$ to access helicity/spin information in jetty cascades
 \item generalizing $\mathcal{G}(R)$ to some kind of $n$-particle correlation function, which might prove to be useful in the context of
 $n$-body decays
 \item using $\Delta \mathcal{G}(R)$ to zoom in on the prominent angular scales within a jet and defining
 some kind of `angular filtering' procedure to improve mass resolution 
  \item using $\mathcal{G}(R)$ to study correlations in the underlying event
  \end{itemize}
 By performing what is essentially an `angular fourier transform' on the constituents of a jet, $\Delta \mathcal{G}(R)$ 
provides a convenient way of accessing angular and mass scales within jets.  These angular and mass scales can be 
used to characterize the substructure of a jet.
Further work will be needed to determine the extent to which the ideas explored in this paper can
be applied more generally.

\Acknowledgements

The authors thank Michael Peskin and Jay Wacker for useful
discussions on jets and perturbative QCD.  The authors would also like to thank JoAnne Hewett,
Michael Peskin, and Jay Wacker for helpful feedback on the manuscript.  
We would like to thank Steve Ellis for suggesting the term `cliffs' where we had 
previously (and confusedly) had `ledges.'
  A.L.~thanks Steve Ellis,
Matt Strassler and Jon Walsh for an introduction to jets and motivation
for studying jet substructure when the field was still in its infancy. 
This work is supported by the US Department of Energy under 
contract DE--AC02--76SF00515.  M.J.~receives partial support from the Stanford
Institute for Theoretical Physics and A.L.~is also supported by
an LHC Theory Initiative Travel Award.

\appendix

\section{Top Quark Decay Kinematics}

If we make some simplifying assumptions about the kinematics
of top quark decays, then we can derive compact formulas for the
angular scales $R_{i*}$ where we expect top jets to have significant
substructure.
To do so we first work in the approximation that both the top and the $W^\pm$ decay
isotropically in their rest frames.  Then working in the limit of large transverse
momenta, we can approximate the typical momentum fractions of the decay
products of the top in the lab frame as
\beq
z_{W1} = z_{W2} = \frac{1}{2} z_{W} = { m_t^2 + m_W^2 \over 4 m_t^2}\simeq 0.30 \quad\quad z_b 
=  { m_t^2 - m_W^2 \over 2 m_t^2} \simeq 0.40
\eeq{momfracWb}
A typical configuration \citep{topdecay} has the decay products approximately distributed along a 
line with
\beq
R_{b1} \leq R_{12} \leq R_{b2} 
\eeq{lineapprox}
Assuming that the decay topology is exactly line-like with
\beq
\quad R_{b2} = R_{12} + R_{b1}
\eeq{lineapprox1}
we can use mass constraints to determine the $R_{i*}$ and $m_{i*}$
\begin{center}
\begin{tabular}{c | c c c}
&$i=1$&$i=2$&$i=3$\\
\hline 
\\
$R_{i*}$ & ${2 m_t^2 \over p_{T }}{m_t - m_W \over m_t^2+ m_W^2}$  & ${2m_t^2 \over p_{T }} {2 m_w  \over m_t^2+m_W^2}$ & ${2 m_t^2 \over p_{T }}{m_t + m_W \over m_t^2+ m_W^2}$ \\
\\
\hline
\\
 $m_{i*}^2$ &  ${(m_t-m_W)^2 \over 2} { m_t^2-m_W^2  \over m_t^2+m_W^2}$&  $m_W^2$  &  ${(m_t+m_W)^2 \over 2} { m_t^2-m_W^2  \over m_t^2+m_W^2}$\\
 \\
\end{tabular}
\end{center}
where $p_T$ is the transverse momentum of the top quark.  Numerical values of these expressions
for $p_T=550$ GeV are given in Sec.~3.2.

\end{document}